\documentclass[fleqn,10pt,twocolumn]{wlscirep}
\usepackage[utf8]{inputenc}
\usepackage[T1]{fontenc}

\newcommand{\orcid}[1]{\href{https://orcid.org/#1}
{\includegraphics[width=10pt]{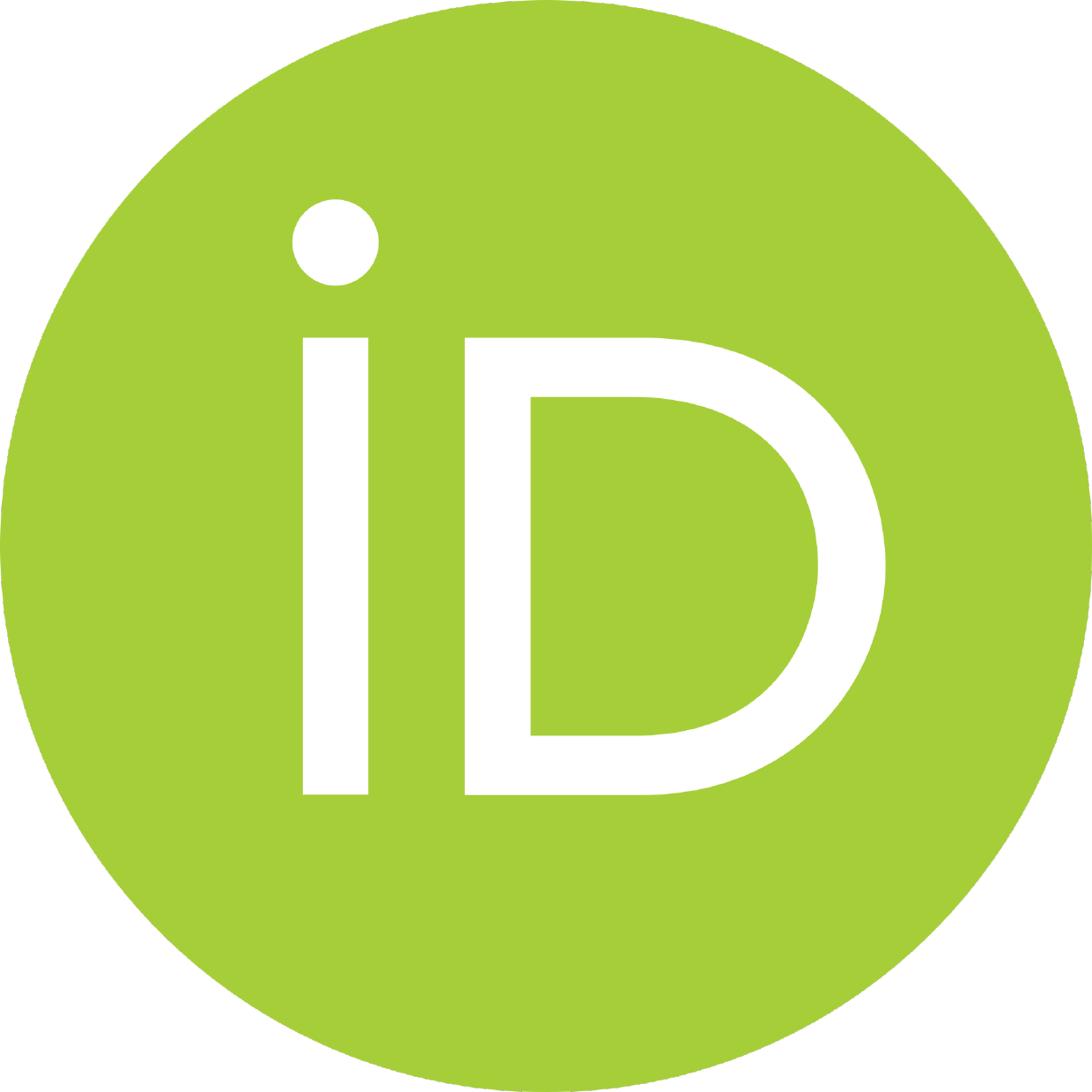}}}

\title{The Diffuse Gamma-Ray Flux from Clusters of Galaxies}

\author[1,2*]{Saqib Hussain}
\author[3,4]{Rafael {Alves Batista}}
\author[1]{Elisabete M. {de Gouveia Dal Pino}}
\author[5,6]{Klaus Dolag}

\affil[1]{Institute of Astronomy, Geophysics and Atmospheric Sciences (IAG), University of S\~ao Paulo (USP), R. do Matão, 1226, 05508-090, S\~ao Paulo, Brazil}
\affil[2]{Gran Sasso Science Institute, Via Michele Iacobucci, 2, 67100 L'Aquila, Italy, Email: saqib.hussain@gssi.it}
\affil[3]{Instituto de F\'isica Te\'orica UAM-CSIC, C/ Nicol\'as Cabrera 13-15, 28049 Madrid, Spain}
\affil[4]{Departamento de F\'isica Te\'orica, Universidad Aut\'onoma de Madrid, M-15, 28049 Madrid, Spain}
\affil[5]{University Observatory Munich, Scheinerstr. 1, 81679 M\"unchen, Germany}
\affil[6]{Max Planck Institute for Astrophysics, Karl-Schwarzschild-Str 1, 85741 Garching, Germany}

\begin{abstract}

The origin of the  diffuse gamma-ray background (DGRB),
the one that remains after subtracting all individual sources from observed gamma-ray sky, is unknown. The DGRB possibly encompasses contributions from different source populations such as star-forming galaxies, starburst galaxies, active galactic nuclei,  gamma-ray bursts, or galaxy clusters.  Here, we combine cosmological magnetohydrodynamical simulations of clusters of galaxies with the propagation of cosmic rays (CRs) using Monte Carlo simulations, in the redshift range $z\leq 5.0$, and show that the integrated gamma-ray flux from clusters can contribute up to $100\%$ of the DGRB flux observed by Fermi-LAT above $100$~GeV, for CRs spectral indices $\alpha = 1.5 - 2.5$ and energy cutoffs  $E_{\text{max}} = 10^{16} - 10^{17}$~eV.
The flux is dominated by clusters with masses  $10^{13} \lesssim M/M_{\odot} \lesssim 10^{15}$ and redshift $ z \lesssim 0.3$. Our results also predict the potential observation of high-energy gamma rays from clusters by experiments like the High Altitude Water Cherenkov (HAWC), the Large High Altitude Air Shower Observatory (LHAASO), and potentially the upcoming Cherenkov Telescope Array (CTA).

\end{abstract}

\begin{document}

\flushbottom
\maketitle

 \section*{Introduction}
The DGRB provides a unique glimpse into the high-energy universe. Its inherent links with high-energy CRs and neutrinos enable investigations of the most powerful cosmic accelerators in the Cosmos.
The observed energy fluxes of these three components are all comparable~\cite{ahlers2018opening, fang2018linking, alves2019open}, suggesting that they may have a common origin. 
Galaxy clusters are believed to be the result of very violent processes such as the accretion and merging of smaller structures into larger ones. These processes can release large amounts of energy (about  $10^{60} - 10^{64} \; \text{erg}$), part of which can accelerate CRs to very-high energies \cite{brunetti2014cosmic,bonafede2021coma, nishiwaki2021particle}.
CRs with $E \lesssim 10^{17} \; \text{eV}$ can be confined within clusters for a time comparable to the age of the universe due to the size of these structures (of the order of  $\text{Mpc}$) and their magnetic-field strength ($B \sim \mu \text{G}$)  \cite{dolag2005constrained, brunetti2014cosmic}.
Therefore, clusters are unique reservoirs of CRs that can produce high-energy photons through collisions with the gas in intracluster medium (ICM), or through processes involving energetic electron–positron pairs produced as secondaries of hadronic and/or leptonic interactions. 
CR interactions with the cosmic microwave background (CMB) and the extragalactic background light (EBL) are also  promising channels for producing high-energy gamma rays, especially for CRs with energies $\gtrsim 10^{18} \; \text{eV}$.

Several analytical and semi-analytical models have been employed to estimate the fluxes of gamma rays and neutrinos stemming from CR interactions in
the ICM \cite{berezinsky1997clusters, colafrancesco1998clusters, rordorf2004diffusion, blasi2007gamma, kotera2009propagation, fang2018linking}, but in all these studies the ICM is 
assumed to have spherically symmetric distributions of magnetic fields and gas.

Here, we explore the production of DGRB by galaxy clusters. We adopt a more rigorous numerical approach, employing cosmological three-dimensional magnetohydrodynamic (3D-MHD) simulations~\cite{dolag2005constrained}, taking into account the non-uniform distributions of the gas density, temperature, and magnetic field, as well as their dependence on the mass and redshift of the clusters. 
We did not make any approximations to constrain the background density, temperature, and magnetic fields of the ICM as they are directly obtained from the simulations.
This extends our previous work in which we employed a similar approach to compute the diffuse neutrino emission from these structures~\cite{hussain2021high}.
Our cosmological simulations indicate that the magnetic field and gas density distributions in massive clusters (with $M \gtrsim 10^{14} \; M_{\odot}$) are larger than in the lower-mass ones, and that massive clusters ($M \gtrsim 10^{15} \; M_{\odot}$) are less abundant at high redshifts \cite{jenkins2001mass, rosati2002evolution, bocquet2016halo, hussain2021high}. 
The neutrino flux from clusters obtained in ref.~\cite{hussain2021high}  is comparable with observations by the IceCube Neutrino Observatory 
~\cite{fang2018linking,abbasi2022searching}. Most of the contribution to the total flux comes from clusters at redshift $z \leq 0.3$ with masses $M \gtrsim 10^{14} M_{\odot}$. 

\section*{Results} 

We inject CRs with minimum energy of $100 \; \text{GeV}$, such that we can study gamma-ray energies down to a few $ 10 \; \text{GeV}$.   
The CRs can escape more easily from the regions with lower densities and magnetic-field strengths in the outskirts of the clusters, which decreases the gamma-ray flux.
In Fig. S4 of the Supplementary Material, we show the gamma-ray flux collected at the edge of individual clusters, produced by CR sources in different locations inside them. We find that the flux is one-order of magnitude larger when the source is located in the central region than in the edge of the cluster. 
For this reason, in order to compute the integrated contribution from all clusters in different redshifts below, we consider only the dominant contribution, i.e. from CR sources in the central region of the clusters. 

\begin{figure} 
\includegraphics[width=\columnwidth]{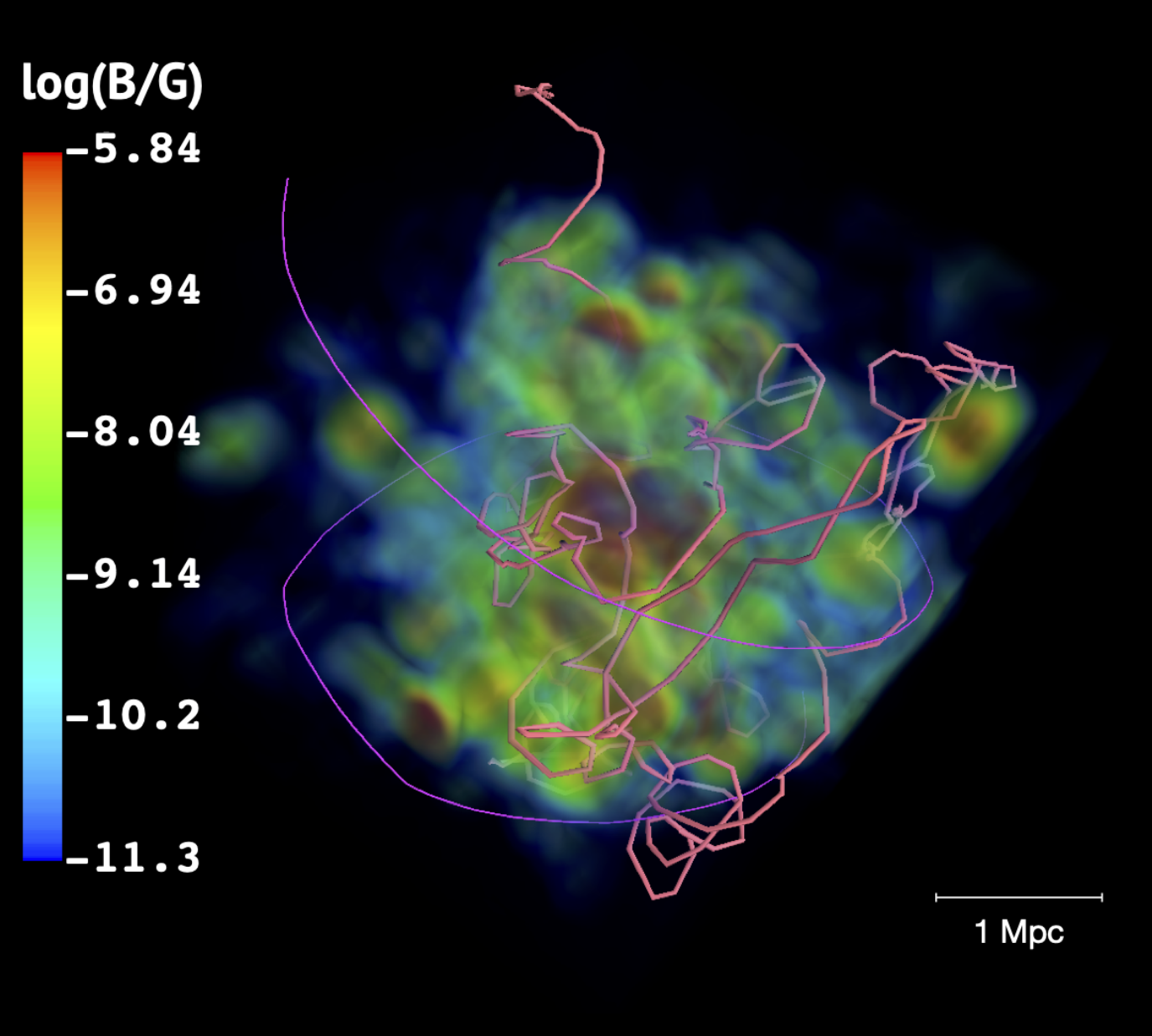} 
\caption{
\textbf{Trajectories of CRs  through a cluster} of mass $\sim 10^{15}\; M_{\odot}$ selected from our  background simulation. The map  depicts the magnetic field intensity distribution in the cluster. The thick (pink) line corresponds to a CR with energy of $10 \; \text{PeV}$, and the thin (purple) line to a CR with energy $500 \; \text{PeV}$. }
\label{fig:trajectory}
\end{figure}

The mass range of clusters in our background simulation is $10^{12} \lesssim M/M_{\odot} < 5\times 10^{15}$ and clusters with masses $\lesssim$ $10^{13} \; M_{\odot}$ barely contribute to the high-energy gamma-ray flux. This occurs due to the lower interaction rate between CRs and the intracluster environment, which is a consequence of the interplay between the Larmor radius, determined by the magnetic field, and the cluster size (see Supplementary Material for a detailed discussion).
Also, massive clusters ($\gtrsim 10^{15} \; M_{\odot}$) exist mostly at low redshifts $z\lesssim 1$, being rare at high redshifts. Therefore, the major contribution to the total flux comes from clusters in the mass range $10^{13} \lesssim M/M_{\odot} \lesssim 10^{15}$ (see Fig. S5).
Fig.~\ref{fig:trajectory} illustrates the propagation of two 
CRs within a cluster of our background simulation.

\begin{figure}
\includegraphics[width=\columnwidth]{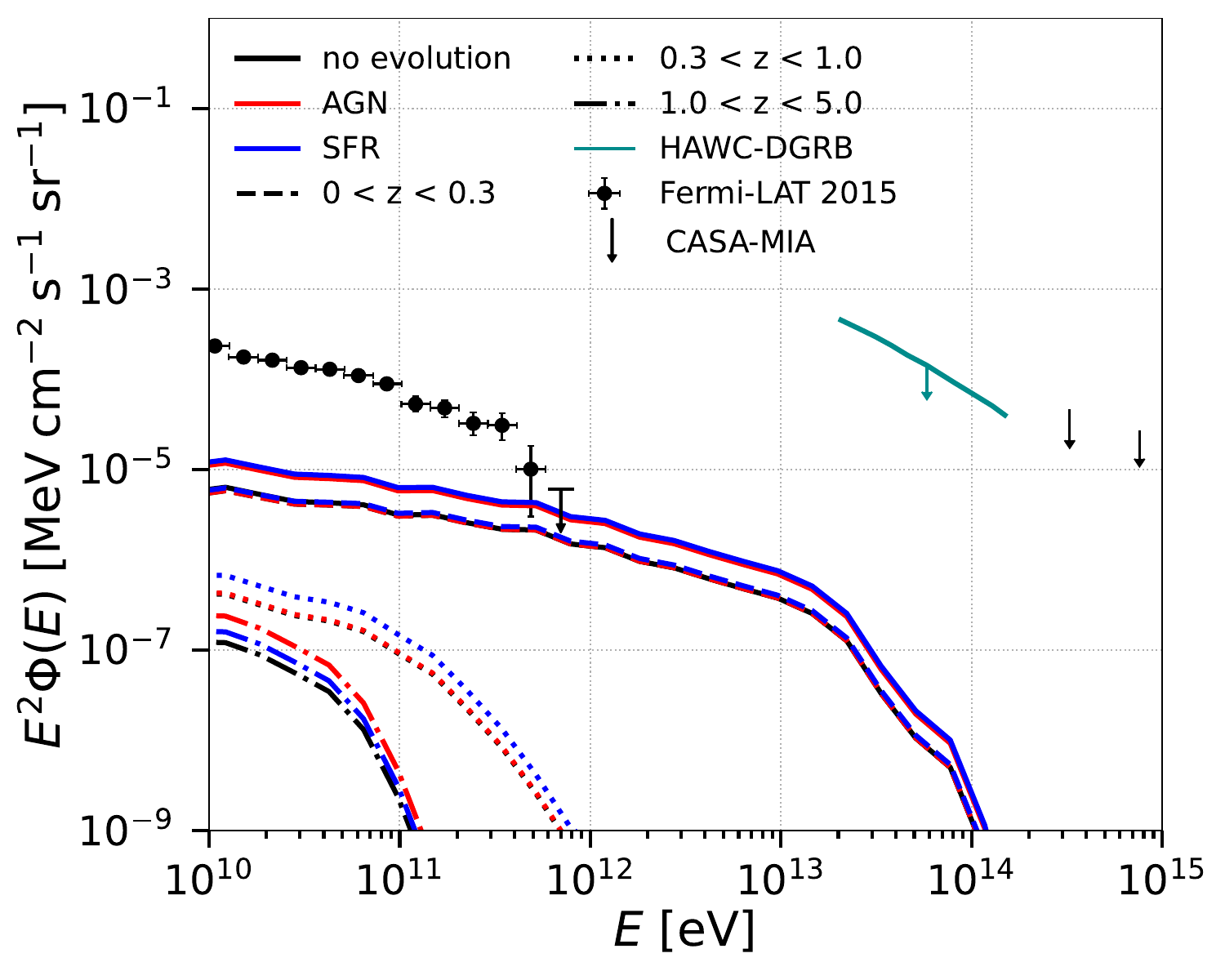}
\caption{\textbf{Gamma-ray flux from clusters at different redshift intervals.} Total flux of gamma rays
for $\alpha =2.3$ and $E_{\text{max}}=10^{17}$~eV over the entire redshift range (solid lines) and also for different redshift intervals (dash-dotted, dotted, and dashed). 
The solid lines are the sum of dashed, dotted and dashed-dotted lines.
The figure also compares the flux including the separated 
contributions  of the evolution of the CR sources (AGN and SFR) with the flux when there is no source evolution. For comparison, the observed flux by Fermi-LAT is depicted (error bars correspond to the total uncertainties, statistical and systematic) \cite{ackermann2015spectrum}, as well as the upper limits  obtained by HAWC ($95\%$ confidence level)~\cite{harding2019constraints}
and CASA-MIA ($90\%$ confidence level)~\cite{chantell1997limits} experiments. 
}\label{fig:ph_UDS-SMP}
\end{figure}

\begin{figure} 
\includegraphics[width=\columnwidth]{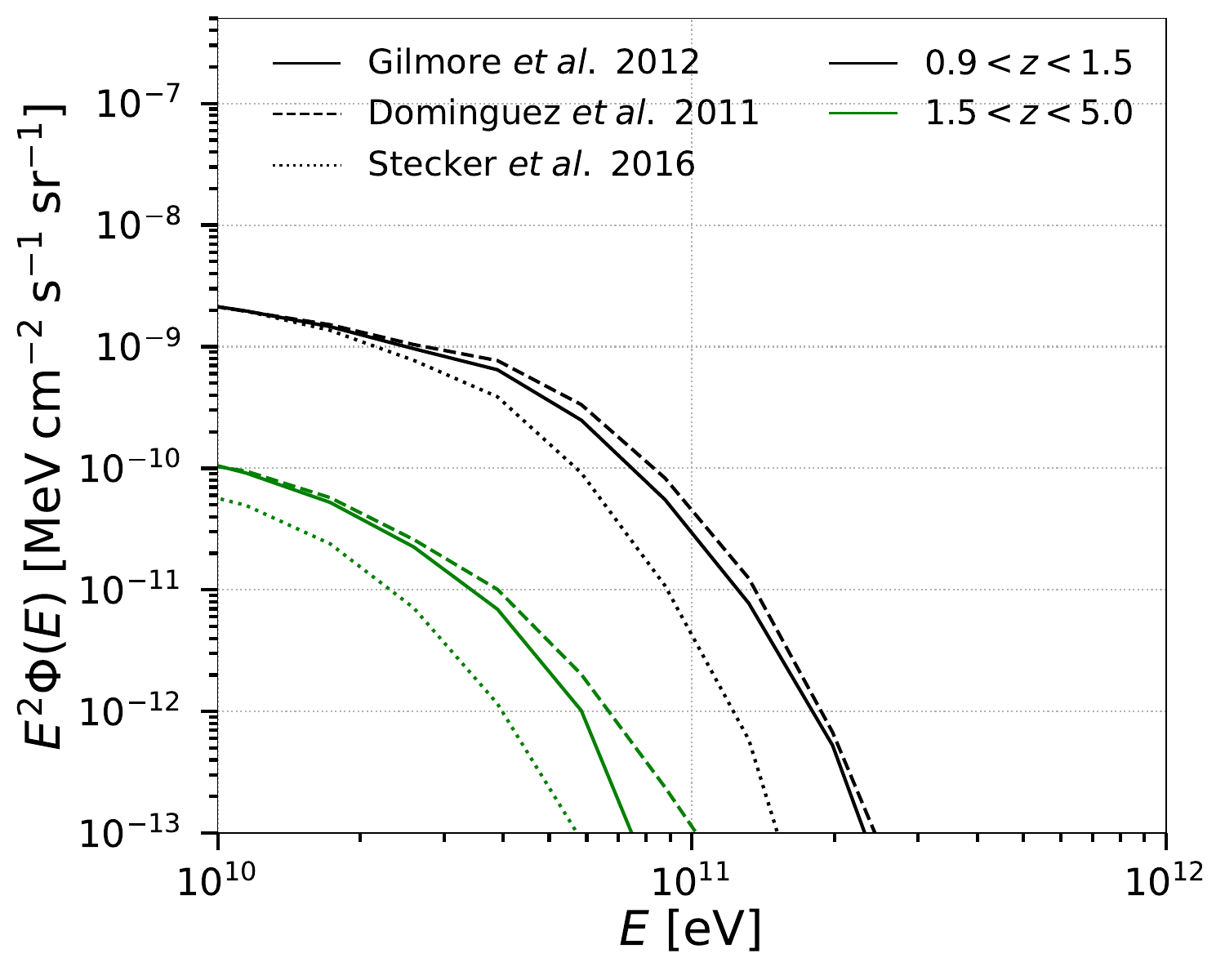} 
\caption{\textbf{Effect of EBL attenuation on the gamma-ray flux} for two redshift intervals and three different EBL models \cite{gilmore2012semi, dominguez2011extragalactic, stecker2016empirical}. 
This figure indicates that the EBL flux attenuation is more prominent at high redshifts and sensitive to the adopted EBL model. The flux is plotted for $\alpha=2.3$ and $E_\text{max}=10^{17}$~eV. }\label{fig:ph_EBLmodels}
\end{figure}

\begin{figure} 
\includegraphics[width=0.495\textwidth]{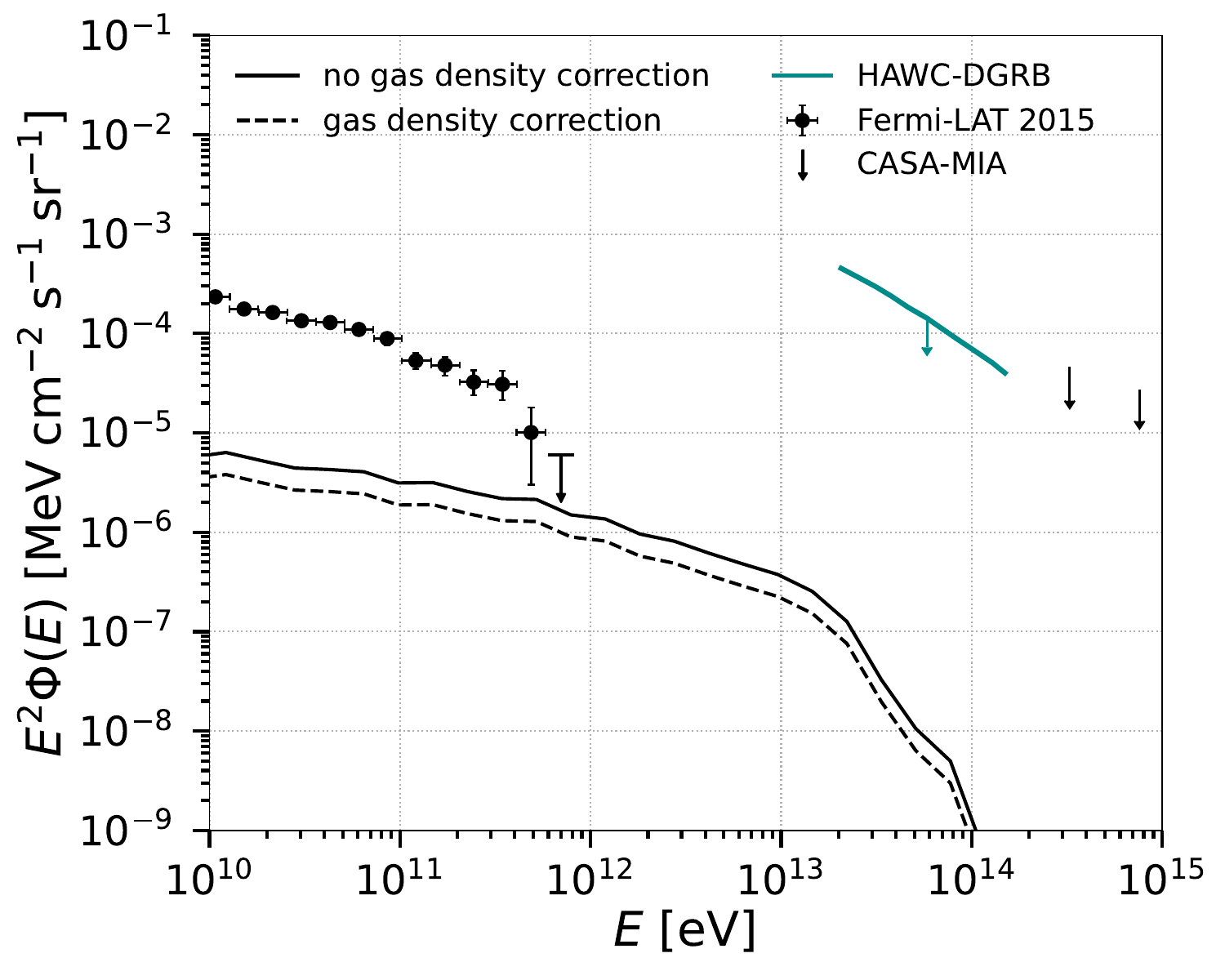}
\caption{
\textbf{Total gamma-ray flux for  $\alpha = 2.3$ and $E_\text{max} = 10^{17}$~eV} over the entire redshift range as, in Fig.~\ref{fig:ph_UDS-SMP} (solid black line). It is compared with the total gamma-ray flux that we obtain when accounting for the gas loss of the clusters due to star formation and AGN feedback (black dashed line). The figure also shows the DGRB  observations from Fermi LAT (error bars correspond to the total uncertainties, statistical and systematic) \cite{ackermann2015spectrum}, as well as the upper limits obtained by HAWC ($95\%$ confidence level)~\cite{harding2019constraints} and  CASA-MIA ($90\%$ confidence level) \cite{chantell1997limits} experiments.
}
\label{fig:phSp-gasCorrection}
\end{figure}

In Figs.~\ref{fig:ph_UDS-SMP} - \ref{fig:ph_BandSensitivity}
we present the integrated gamma-ray spectrum from all clusters for $z \leq 5.0$, propagated up to the Earth. The total flux ($\Phi$) was computed as follows:
\begin{align} 
E_\text{obs}^{2} \Phi(E_\text{obs}) &=  \int\limits_{z_\text{min}}^{z_\text{max}} \text{d}z \int\limits_{M_\text{min}}^{M_\text{max}} \text{d}M \dfrac{\text{d}N}{\text{d}M} E^{2}  \dfrac{\text{d}\dot{N}( E/(1+z), M , z)}{\text{d}E} \nonumber \\ 
&  g(E_\text{obs}, E, z)
\left( \dfrac{\psi_{\text{ev}}(z) f(M)}{4\pi d_L^2 (z)}\right)
\end{align}
where the number of clusters per mass interval $dN/dM$ was calculated from our background simulation (see Fig.~S1), $g(E_\text{obs}, E, z)$
accounts for the interactions of gamma rays with energy $E$ arriving with energy $E_\text{obs}$ undergoing interactions during their propagation in the ICM and the intergalactic medium (IGM),
$\psi_{\text{ev}}(z)$ is a function that describes the cosmological evolution of the emissivity of the CR sources (AGN, SFR, or none; see Equations E1 and E2 of the Supplementary Material), the quantity $E^2 \; d\dot{N}/dE$ denotes the gamma-ray power computed from the simulation,  $d_L$ is the luminosity distance, and $f(M)$ is a factor of order unit that corrects the flux by the amount of gas that is removed from the clusters due to stellar and AGN feedback.
We note that the number of clusters per mass interval we obtained from our MHD cosmological simulation at different redshifts is comparable with results from other large-scale cosmological simulations\cite{jenkins2001mass, rosati2002evolution, bocquet2016halo} and predictions from observations \cite{giovannini1999radio, tinker2008toward} (see  Fig.~S1).

The universe is believed to be isotropic and homogeneous at very large scales. Therefore, for the propagation of gamma rays from the clusters to Earth, we assumed a nearly uniform distribution of sources in comoving coordinates.

Fig.~\ref{fig:ph_UDS-SMP} depicts the total flux for different redshift intervals: $z \leq 0.3$, $0.3 < z \leq 1.0$, and $1.0 < z \leq 5.0$. 
A representative  spectral index $\alpha=2.3$ and a maximum energy $E_\text{max} = 10^{17}$ eV are used for this evaluation 
(see also Figs.~\ref{fig:ph_EBLmodels}-\ref{fig:phSp-gasCorrection}).
The dominant contribution to the total flux of gamma rays comes from sources at low redshifts ($z \lesssim 0.3$), for which the effect of the EBL attenuation is less pronounced. 
This effect is more prominent at higher redshifts and also depends on the EBL model adopted~\cite{gilmore2012semi, dominguez2011extragalactic, stecker2016empirical} (see  Fig.~\ref{fig:ph_EBLmodels}, and Fig.~S7 of the Supplementary Material). Fig.~\ref{fig:ph_UDS-SMP} shows the results for the EBL model from ref.~\cite{gilmore2012semi}, which predicts a slightly larger gamma-ray cut-off energy for the flux. 
Also, our treatment of the pp-interactions~\cite{kafexhiu2014parametrization, kelner2006energy} is only an approximation and contains uncertainties due to the unknown pp cross-section at energies beyond the reach of the LHC~ \cite{aaboud2016measurement}.

Fig.~\ref{fig:ph_UDS-SMP} also highlights the effects of the evolution of the CR sources on the gamma-ray flux, distinguishing the separated  contributions of AGN and SFR, following the same procedure as in refs. \cite{batista2019cosmogenic, hussain2021high}. 
We find that an AGN-type evolution enhances the diffuse gamma-ray flux at high redshifts ($z \gtrsim 1.5$) compared to scenarios wherein the sources evolve as the SFR (or without any evolution). 
On the other hand, these contributions are both comparable at low redshifts ($z \lesssim 0.3$) which in turn, provide the dominant contribution to the total gamma-ray flux.

We further notice that the flux of gamma rays above energies $\sim 10^{12}$~eV can also be attenuated by interactions with the local optical and infrared photon fields of clusters, in addition to the EBL. Nevertheless, this effect is more dominant for sources at redshift $z \gtrsim 0.3$ as discussed in ref.~\cite{murase2016constraining}. In our case, the major contribution corresponds to sources at $z \lesssim 0.3$. Therefore, we expect that this interaction channel has likely a minor impact on our results.

As remarked, our MHD simulations do not include radiative-cooling, or the amount of gas that is converted into stars or removed from the clusters due to stellar and AGN feedback. This implies a slight overestimation of the density in the structures, especially for clusters of mass $\lesssim 10^{14}\; M_{\odot}$ (see refs. \cite{fabjan2010simulating,planelles2014role}). Based on observational results~\cite{lovisari2015scaling}, we have also estimated the total gamma-ray flux taking into account the expected decrease of the gas density as a function of the cluster mass. 
In Fig.~\ref{fig:phSp-gasCorrection} we recalculated the total diffuse gamma-ray flux (black dashed line) considering correction factors  $f(M) \sim 0.95$  for clusters with $M \gtrsim 10^{15}\; M_{\odot}$, $f(M) \sim 0.8$ for $M\gtrsim 10^{14}\; M_{\odot}$, $f(M) \sim 0.3$ for $M\gtrsim 10^{13}\; M_{\odot}$,  and $f(M) \sim 0.3$ for $M\gtrsim 10^{12}\; M_{\odot}$, following ref.~\cite{lovisari2015scaling}.
A comparison between the dashed and solid black lines of Fig.~\ref{fig:phSp-gasCorrection} indicates a small reduction of the flux by at most a factor about $ 2$.

The results for different combinations of the CR cutoff energy  and spectral index  are presented in Fig.~\ref{fig:ph_BandSensitivity}.
The shaded region shows the total flux of gamma rays for all clusters from the entire redshift range $0<z \leq 5.0$, calculated for  $\alpha = 1.5-2.5$ and  $E_{\text{max}} = 10^{16} - 10^{17} \; \text{eV}$, including feedback by AGN and SF, and CR source evolution.
The observed DGRB flux by Fermi-LAT, and the upper limits  obtained by the currently operating HAWC~\cite{harding2019constraints} and by the CASA-MIA experiment~\cite{chantell1997limits}, are also shown.
For energies greater than $\gtrsim 100 \; \text{GeV}$, our simulations indicate that galaxy clusters can contribute substantially to the DGRB measured or constrained by these experiments. 
This contribution amounts for up to 100\% of the observed flux by Fermi-LAT, for spectral indices $\alpha \lesssim 2$ and maximum energy $E_\text{max} \gtrsim 10^{17}$~eV. This also clearly explains the apparent flatness of the spectrum up to about $1$~TeV (see also Figs. S8 and S9 of the Supplementary Material).

\subsection*{Discussion}
The spectral indices considered here are consistent with the universal CR model~\cite{pinzke2010simulating}
used by Fermi-LAT to explore the CR induced gamma-ray emission from clusters~\cite{ackermann2014search}, and by H.E.S.S. for the Coma cluster ($\alpha = 2.1 - 2.4$)~\cite{aharonian2009constraints}, while the $E_{\text{max}}$ range is compatible with the fact that the clusters can confine mainly CRs with energies  $E\lesssim 10^{17}$~eV~\cite{fang2016high, hussain2021high}.

Note that the slope of the integrated gamma-ray flux is  strongly influenced by the spectral parameters of the injected CRs. Therefore, when considering potential values for the CR spectral index, it is also important to discuss the corresponding particle acceleration mechanism(s). If CRs are accelerated by the same processes that produce ultra-high-energy CRs (UHECRs), phenomenological fits~\cite{auger2017combined,batista2019cosmogenic, heinze2019new} to the UHECR data favor very hard spectra, with possible spectral indices extending as low as $\alpha < 0$ in some cases (for $E^{-\alpha}$). Such scenarios 
might seem surprising, at first, 
but there are sound explanations that include magnetic field effects, plasma instabilities, re-acceleration, magnetic reconnection, interactions in the sources, etc (see, e.g., \cite{brunetti2014cosmic, brunetti2017relativistic, alves2019open, medina2021particle} for an overview on some of these mechanisms). Naturally, the spectral properties of CRs injected in the ICM in the PeV-EeV range do not need to be the same as the UHECRs, but 
it is reasonable to expect a connection. 
Therefore, even hard spectral indices are theoretically possible. Note that if the CRs responsible for producing the gamma rays are accelerated not by CR sources embedded in clusters but via accretion or merger shocks, for example, then softer spectra ($\alpha \sim 2.0 - 2.3$) are expected.
In Fig.~S8 of the Supplement Material, we show the gamma-ray flux for different combinations of the parameters $\alpha \; \text{and} \; E_{\text{max}}$.

\begin{figure}
\includegraphics[width=0.495\textwidth]{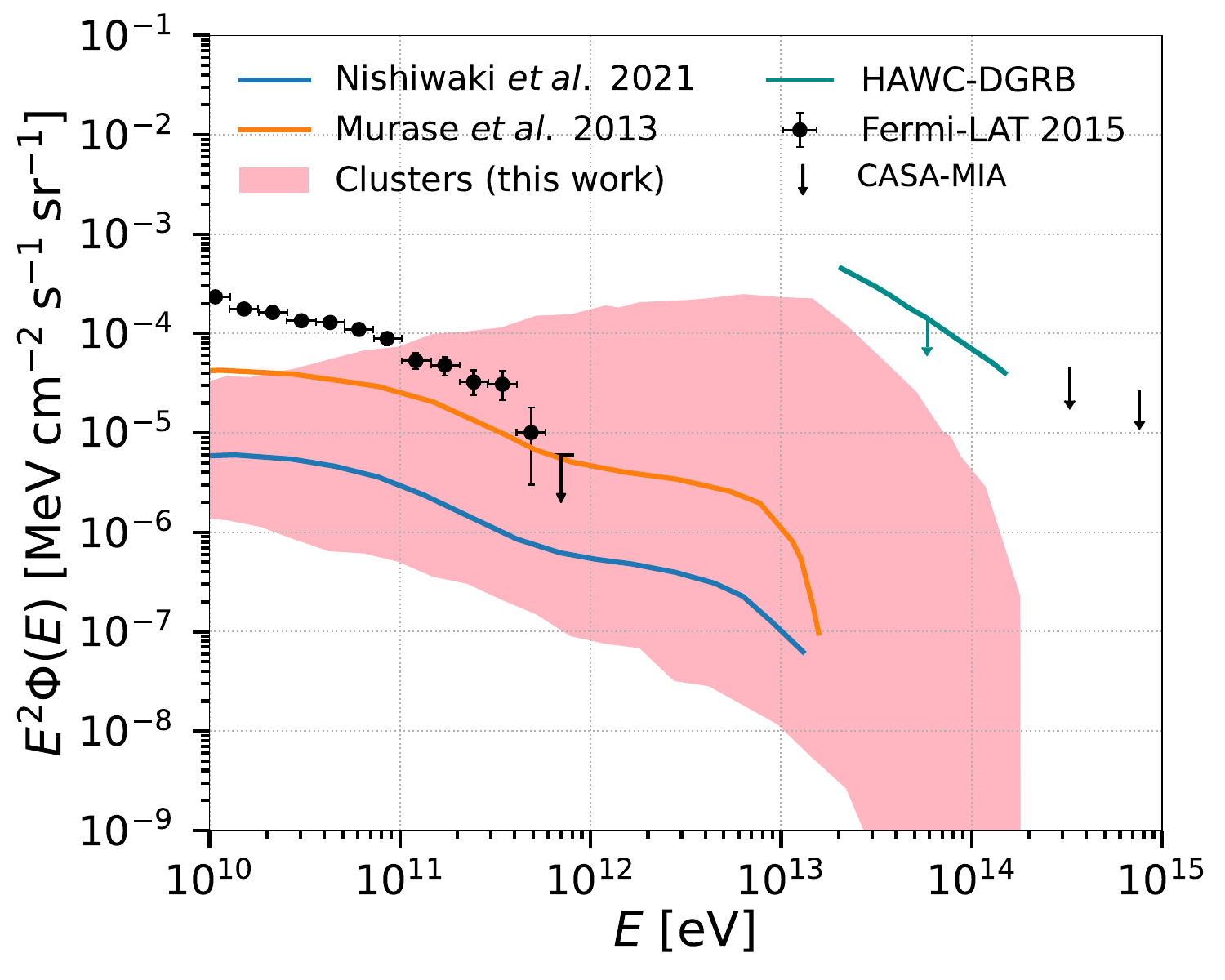}
\caption{\textbf{Integrated gamma-ray flux from the entire population of clusters}. The pink shaded region represents the integrated gamma-ray flux obtained in this work for $E_\text{max} = 10^{16} - 10^{17}$ eV and  spectral index $\alpha= 1.5 - 2.5$, as well as all source evolutions considered (AGN, SFR, and no evolution). This is compared  with the total gamma-ray flux from clusters obtained in previous works  \cite{murase2013testing,nishiwaki2021particle}, and also  with the DGRB  observations from Fermi-LAT (error bars correspond to the total uncertainties, statistical and systematic) \cite{ackermann2015spectrum}, as well as the upper limits obtained by HAWC ($95\%$ confidence level)~\cite{harding2019constraints} and  CASA-MIA ($90\%$ confidence level) \cite{chantell1997limits} experiments.  
}\label{fig:ph_BandSensitivity}
\end{figure}

In Fig.~\ref{fig:ph_BandSensitivity},  the gamma-ray flux we obtained from the entire population of clusters 
is also compared  with  the expected one from Coma-like clusters~\cite{nishiwaki2021particle}, and that obtained in ref.~\cite{murase2013testing}.
 In ref.~\cite{murase2013testing}, they estimated the gamma-ray flux from clusters using a purely hadronuclear scenario ($pp$-interaction)  claiming that these sources would contribute to the DGRB with at least $30\% - 40\%$, or even $100\%$ if the spectrum is soft ($\alpha \gtrsim 2.2$).
In comparison with the estimated spectrum for Coma-like clusters\cite{nishiwaki2021particle}, our gamma-ray flux is a little higher.
In both studies \cite{murase2013testing, nishiwaki2021particle}, besides the oversimplified ICM magnetic-field and density distributions, assumed to have radial profiles, they did not account for the contributions from clusters of mass $\lesssim 10^{14} \; M_{\odot}$.
In Coma-like clusters~\cite{nishiwaki2021particle}, where masses are  of the order of  $10^{15}\; M_{\odot}$, the average density is $\sim 10^{-6} \; \text{Mpc}^{-3}$, but it is $\sim 10^{-4} \; \text{Mpc}^{-3}$ for cluster masses of a few $10^{14}\; M_{\odot}$ (as considered in refs. \cite{nishiwaki2021particle, murase2013testing}), and can be even larger for masses $< 10^{14} \; M_{\odot}$,
as predicted by large scale cosmological simulations \cite{jenkins2001mass, rosati2002evolution, bocquet2016halo} and obtained in our own MHD simulations.
Because we are considering here the entire mass range ($10^{12} \leq M/M_{\odot} < 5\times 10^{15}$), the  density is higher by an order of magnitude, and this is the main difference between ours and these previous studies~\cite{murase2013testing, nishiwaki2021particle}.

Another study~\cite{zandanel2015high} estimated the flux using a simple relation between the gamma-ray luminosity and the cluster mass. 
They constrained the radio-loud cluster count from observations  by the Radio Astronomy Observatory Very Large Array sky survey \cite{giovannini1999radio, cassano2010connection}
and also assumed that the radio luminosity scales linearly with the hadronic high-energy emission. Their results are also comparable with ours.

Though individual source populations such as blazars \cite{ajello2015origin, Raniere2022isotropic}, misaligned-AGNs \cite{di2013diffuse} and star-forming galaxies (SFGs) \cite{roth2021diffuse} can contribute to a fairly large fraction to the DGRB for energies below TeV \cite{fornasa2015nature, ackermann2016resolving} (see Fig. S9 of the Supplementary Material),
our results demonstrate that the cumulative gamma-ray flux from clusters can dominate over the integrated contribution of individual classes of unresolved sources, at energies $\gtrsim 100$~GeV. The implications of our calculations are extremely important considering that the contribution from clusters is guaranteed if high-energy CRs are present in the ICM.

As shown in Fig.~\ref{fig:ph_BandSensitivity}, our results are compatible with  upper limits evaluated by HAWC~\cite{harding2019constraints}. A similar estimate has yet to be performed by other facilities like the LHAASO~\cite{wang2022extra} or the forthcoming CTA~\cite{cta2018science}. Nevertheless, considering the sensitivity curves for point sources obtained in both cases \cite{wang2022extra, cta2018science}, the gamma-ray flux we derived has likely the potential to be detected by these  facilities too (see also Fig. S9 and the discussion therein).

Future more realistic MHD cosmological simulations that account directly for the CR sources distribution, evolution,  and feedback  \cite{barai2019intermediate, hopkins2021first} may allow to constrain better the contribution of clusters to the DGRB. 
Furthermore, the effects of unknown magnetic fields of the diffuse IGM on the gamma-ray cascading may also change our results (see discussion in the Supplementary Material).

Fig.~\ref{fig:multi-messenger_cluster} summarizes our findings, together with those from ref.~\cite{hussain2021high}. It shows both the high-energy gamma-ray and neutrino emission from the entire population of clusters up to redshift $z\leq 5.0$, assuming CR sources embedded in clusters. As we see, the neutrino flux we obtain is comparable with the diffuse neutrino background observed by IceCube for CR spectral index $\alpha = 1.5 - 2.5$ and maximum energy $10^{16} - 10^{17}$ eV. A recent analysis by the IceCube Collaboration~\cite{abbasi2022searching} found that less than $\sim 77\%$ of the total diffuse neutrino flux could be due to clusters. While this could, at first glance, seem in conflict with our results, we note that changes in the parameters of our analysis such as the total CR luminosity or the distribution of CR sources within the cluster could 
reduce our estimate. The same is true for the DGRB predictions. Therefore, the link established by Fig.~\ref{fig:multi-messenger_cluster} between the diffuse gamma-ray and the diffuse neutrino backgrounds, should be interpreted minding these caveats.

Our results were obtained through the most detailed simulations to date of three-dimensional particle transport in cosmological environments. Combined with the other known components of the DGRB, our results strongly constrain the fraction of the diffuse flux that could be ascribed to unknown components such as the elusive dark matter. Moreover, it establishes a clear connection between the fluxes of two messengers, neutrinos and gamma rays, which, combined, enables us to indirectly study CRs in clusters even if they are not directly observable.

\begin{figure*}[htb!]
\centering
\includegraphics[width = 1.0\textwidth]{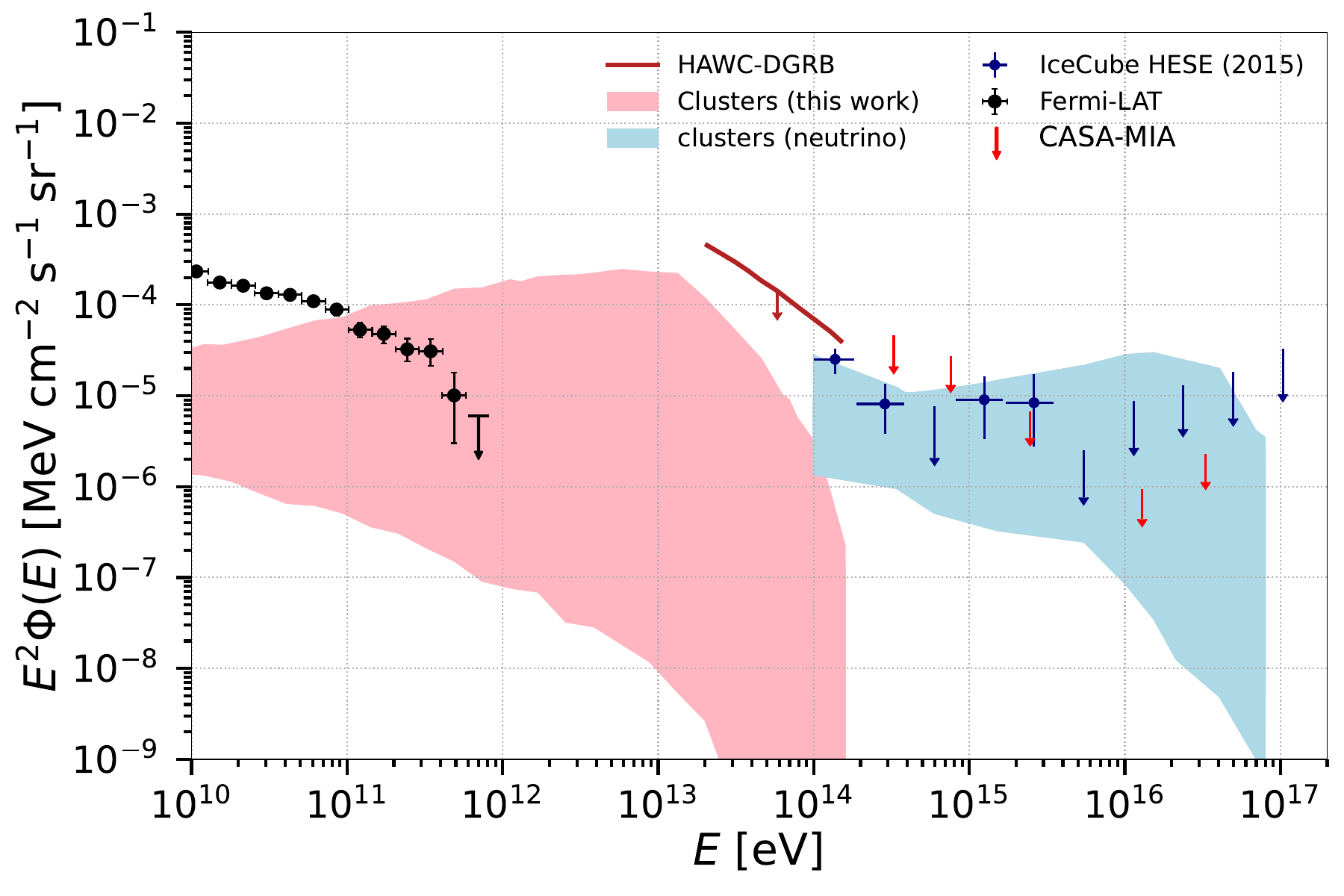}
\caption{\textbf{Multi-messenger emission from clusters of galaxies.}
High-energy neutrinos (blue band) (obtained by \cite{hussain2021high}, error bars in IceCube data  correspond to the $68\%$ confidence intervals \cite{aartsen2015evidence}) and gamma rays (pink band) from the entire population of galaxy clusters obtained in this work. The gamma-ray flux is compared with the DGRB observed by Fermi-LAT (error bars correspond to the total uncertainties, statistical and systematic)~\cite{ackermann2015spectrum}, and the upper limits by HAWC ($95\%$ confidence level) \cite{harding2019constraints} and CASA-MIA ($90\%$ confidence level)~\cite{chantell1997limits}.
}
\label{fig:multi-messenger_cluster}
\end{figure*}

\section*{Methods}\label{sec:simulation-setup}

We describe the ICM through 3D-MHD smoothed-particle-hydrodynamical (SPH) cosmological simulations employing the GADGET code~\cite{springel2001gadget, springel2005cosmological}, within a sphere of radius $110$~Mpc around the Milky Way~\cite{dolag2005constrained}. The simulations extend up to a redshift of $z \simeq 5$ and contain clusters with masses $10^{12} < M/M_{\odot} < 10^{15.5}$.  We consider here seven snapshots at redshifts $0$, $0.05$, $0.2$, $0.5$, $0.9$, $1.5$, and $5.0$.
For clusters in this mass range, the corresponding luminosity interval is about $(10^{42} - 10^{46}) \; \text{erg}\, \text{s}^{-1}$~\cite{schneider2014extragalactic}. These values are used to compute the gamma-ray fluxes shown throughout this work. The  magnetic-field strength varies between  $ 10^{-11} \; \text{G}$ and  $10^{-5} \; \text{G}$ approximately,  which is in reasonable agreement with the expected field strengths from observations of different clusters of galaxies~\cite{bohringer2016cosmic}.
Feedback from active galactic nuclei (AGN) and star formation (SF) are not directly included in these MHD cosmological simulations, but the evolution effects of these potential CR sources on the flux of gamma rays is accounted for with a redshift-dependent profile, as in refs. \cite{hussain2021high,heinze2016cosmogenic, batista2019cosmogenic}
(see equations E1 and E2 of the Supplementary Material).  A flat $\Lambda$CDM universe model is assumed, with the corresponding cosmological parameters given by $h \equiv H_{0}/(100 \; \text{km~s}^{-1}~\text{Mpc}^{-1}) = 0.7$, $\Omega_m = 0.3$, $\Omega_{\Lambda} =0.7$, and the baryonic fraction $\Omega_b/\Omega_m = 14\%$. The maximum resolution in our SPH simulations is approximately $10 \; \text{kpc}$ (see refs.~\cite{hussain2021high, dolag2005constrained} and also page~1 of the Supplementary Material for details).


We are interested in high-energy gamma rays with $E\gtrsim 10 \; \text{GeV}$ whose origin is more uncertain (see Fig.~S9 of the Supplementary Material) and thus  consider CRs with energies $10^{11} \leq E/\text{eV} \leq 10^{19}$. 
The energy around  $10^{19}$~eV can be achieved by primary sources inside a cluster, such as AGNs~\cite{fang2018linking, murase2008cosmic}. For magnetic fields of $B \sim 1 \; \mu\text{G}$, the Larmor radius of CRs with $E \sim 10^{19} \; \text{eV}$ is $r_{\text{L}} \sim 10 \; \text{kpc}$, so that they cannot remain  trapped within clusters for too long.
On the other hand, CRs with lower energies remain confined,  producing secondaries due to interactions with the ICM gas and the bremsstrahlung radiation, as well as  with the CMB and the EBL~\cite{pfrommer2008simulating, murase2008cosmic, fang2016high, hussain2021high, nishiwaki2021particle}.


We explore the propagation of CRs in the simulated background of clusters using the CRPropa code~\cite{batista2016crpropa, alvesbatista2022crpropa}. 
The propagation has two steps and we assume that the CRs are predominantly composed by protons,
since we expect much smaller contribution from heavier elements \cite{kotera2009propagation}  (see page~4 of Supplementary Material).
In the first step, we compute the gamma-ray flux produced by CR interactions in the clusters by considering all relevant interactions that generate both electrons and photons, namely: photopion production, Bethe-Heitler pair production, pair production, inverse Compton scattering, and proton-proton (pp) interactions. In addition, we take into account the energy losses due to the adiabatic expansion of the universe and due to synchrotron emission, although these only contribute to the electromagnetic flux at energies much lower than our energy of interest ($E \gtrsim 10 \; \text{GeV})$. For more details on how CRs were propagated, see the Supplementary Materials (Figs. S2 and S3).
We find that the interactions of the CRs with the cluster gas and the CMB are the dominant channels for producing the secondaries \cite{hussain2021high}.
In the second step, we perform the propagation of the gamma rays collected at the boundary of the clusters to Earth. We consider the electromagnetic cascade process initiated by these gamma rays both in the ICM and in the intergalactic medium, including inverse Compton scattering, single, double, and triple pair production, with the CMB, the EBL~\cite{gilmore2012semi}, and the radio background~\cite{protheroe1996new} (see Fig. S3 in Supplementary Material).
We did not consider the effects of intergalactic magnetic fields outside the cluster in this step, since they are highly uncertain~\cite{vazza2017simulations} and are not expected to majorly affect the gamma-ray flux at energies above $100 \; \text{GeV}$~\cite{alves2021gamma}. 

To compute the gamma-ray flux we have followed the same procedure given in ref.~\cite{hussain2021high} and considered that $1\%$ of the cluster luminosity goes into CRs, which is consistent with Fermi-LAT predictions~\cite{ackermann2014search}. 
We only considered the contribution of CRs with energies above  $100 \; \text{GeV}$ approximately, although we did consider the whole energy range, starting from 1 GeV, to normalize the total energy of the simulation to the cluster luminosity, as explained on page~4 
in the Supplementary Material. 
Also, in Fig. S6 of the Supplementary Material, we compare the gamma-ray flux for different values of this luminosity fraction and the results indicate a variation much less than an order of magnitude.

\section*{Data availability}
The datasets generated during and/or analysed during the current study are available from the corresponding author upon request.

\section*{Code availability}
The numerical codes used to generate results that are reported in the current study are available  upon request.



\section*{Acknowledgements}
SH  
acknowledges support from the Brazilian funding agencies CNPq  and FAPESP (grant 2013/10559-5). RAB is funded by  ``la Caixa'' Foundation (ID 100010434) and  European Union's Horizon~2020 research and innovation program under the Marie Skłodowska-Curie grant agreement No 847648  (fellowship  LCF/BQ/PI21/11830030);  he was also supported by the grants PID2021-125331NB-I00 and CEX2020-001007-S funded by MCIN/AEI/10.13039/501100011033 and by ``ERDF A way of making Europe''. 
EMdGDP also acknowledges support from the Brazilian agencies FAPESP (grant 2013/10559-5) and CNPq (grant 308643/2017-8). 
KD  acknowledges support by Deutsche Forschungsgemeinschaft (DFG, German Research Foundation) under Germany’s Excellence Strategy – EXC-2094 – 390783311 and by  funding for the COMPLEX project from the European Research Council (ERC) under the European Union’s Horizon 2020 research and innovation program (grant  ERC-2019-AdG 882679).
The numerical simulations presented here were performed in the cluster 
GAPAE of IAG/USP 
(FAPESP grant 2013/10559-5). 

\section*{Author contributions}
SH performed the calculations, analysis and writing of the manuscript. RAB conceived the work, helped with the Monte Carlo simulations, performed analysis and writing. EMGDP coordinated the project and performed analysis and writing. KD performed the MHD simulations. 

\section*{Competing interests}
The authors declare no competing interests.



\appendix

\section*{Supplementary Material}

\bigskip

\begin{figure}[htb!]
	\centering	\includegraphics[width=0.9\columnwidth]{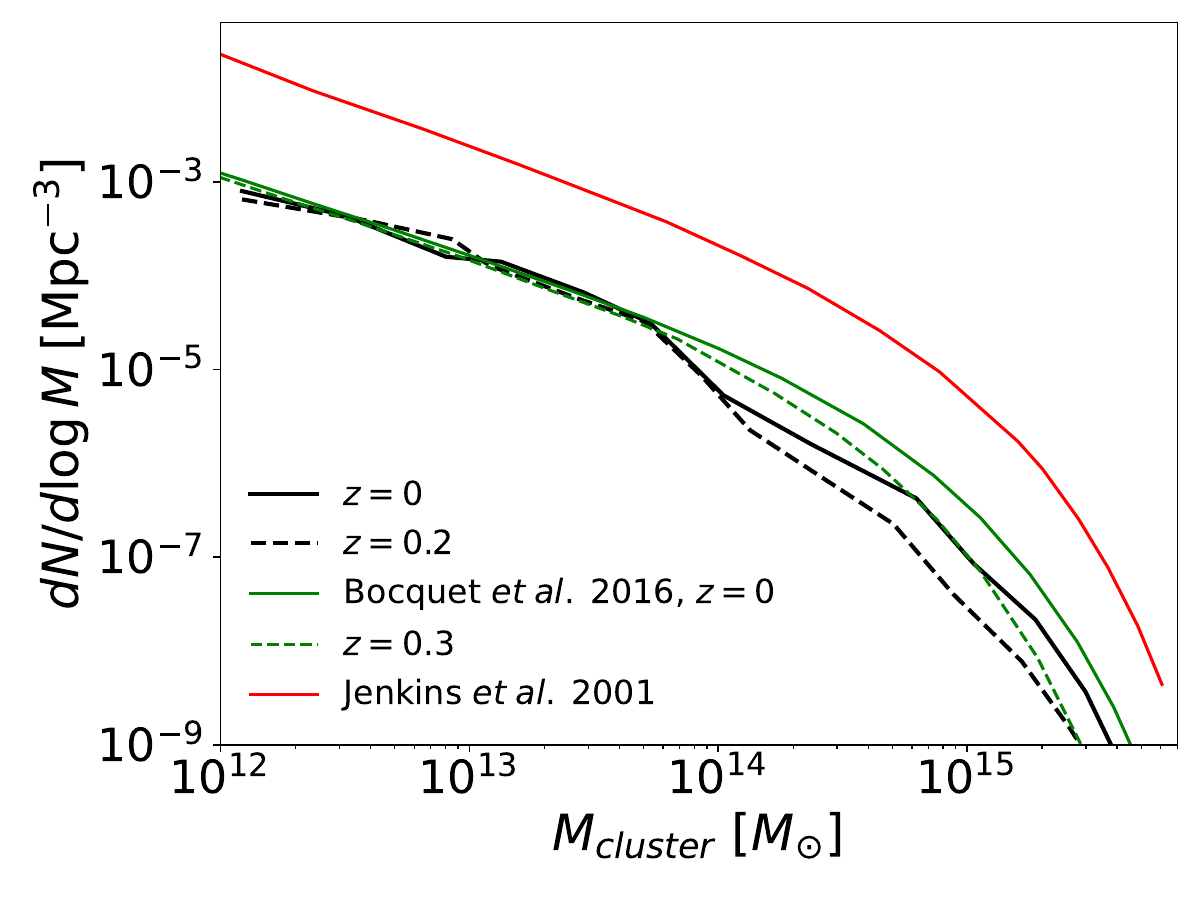}
	\caption{\textbf{Cluster density:}
 Black lines represent the number of clusters per mass interval in our cosmological simulation for different redshifts. For comparison, results from other large-scale cosmological simulations are also shown as green~\cite{bocquet2016halo}and  red lines~\cite{jenkins2001mass}. We note that in ref.~\cite{jenkins2001mass} (red line) it is presented the total count of clusters as a function of mass starting at redshift $z=14$ up to $z = 0$. This explains the difference with regard to the other curves.}
	\label{fig:ClusterMassFunc}
\end{figure}

\paragraph{Cosmological simulations.} 
To calculate the contribution of galaxy clusters to the diffuse gamma-ray background (DGRB), we employed three-dimensional cosmological MHD simulations~\cite{dolag2005constrained} obtained with the GADGET code~\cite{springel2001gadget,springel2005cosmological}. They cover a large volume (a sphere of radius $\sim 110 \; \text{Mpc})$ and a redshift range  $z \simeq 0-5$, and contain several clusters with masses $10^{12} < M/M_{\odot} < 10^{15.5}$.  At $z=0$, these simulations reproduce quite well the distribution of nearby clusters, including  Virgo, Perseus, and Coma, which are within  $100$~Mpc away from Earth approximately. 
The number of clusters per mass interval we obtained  at different redshifts is comparable with results from other large-scale cosmological simulations \cite{jenkins2001mass, rosati2002evolution, bocquet2016halo} (see Fig.~\ref{fig:ClusterMassFunc}), and with predictions from observations~\cite{giovannini1999radio, tinker2008toward}.

\begin{figure}[htb]
\centering
\includegraphics[width=0.495\textwidth]{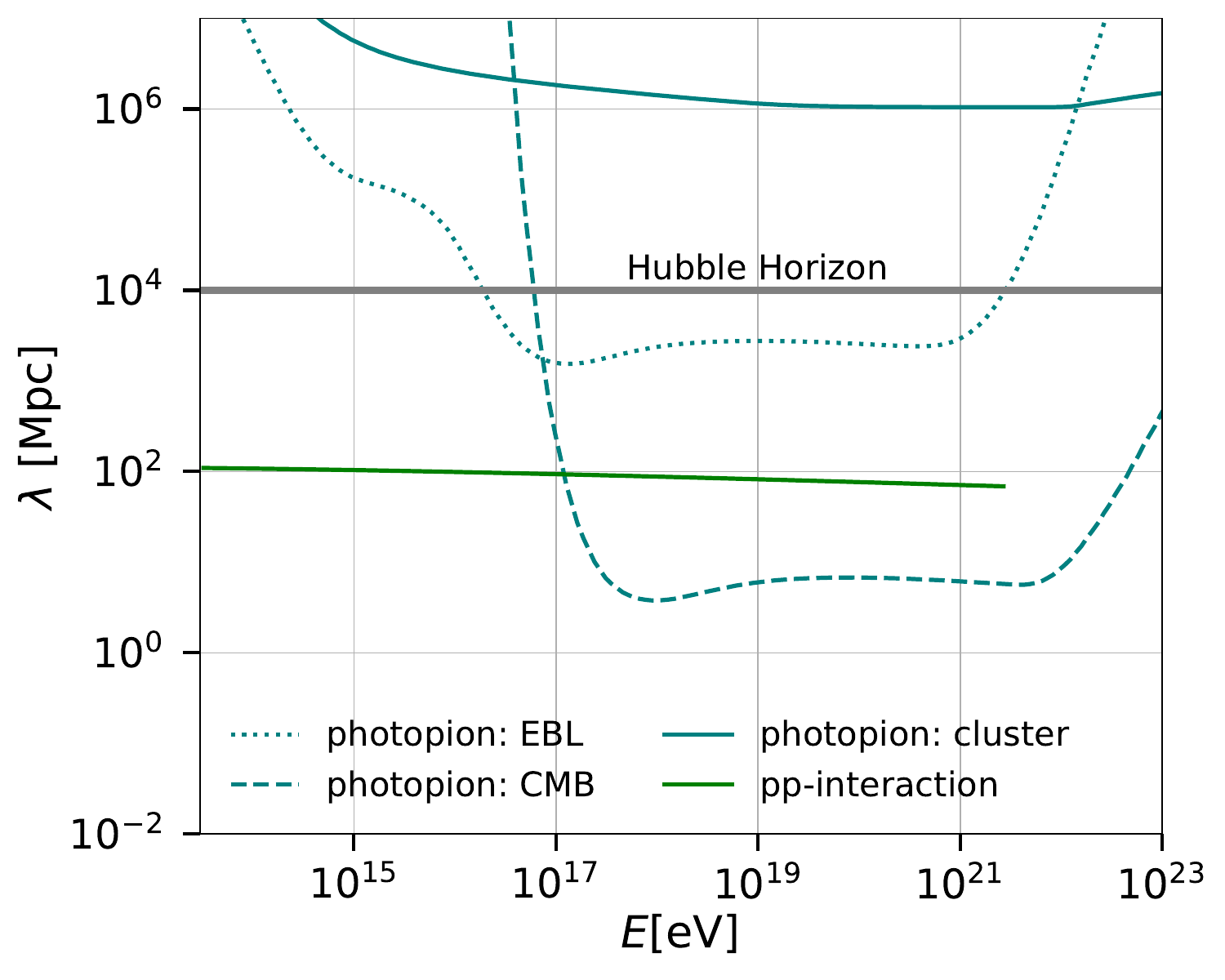}
\includegraphics[width=0.495\textwidth]{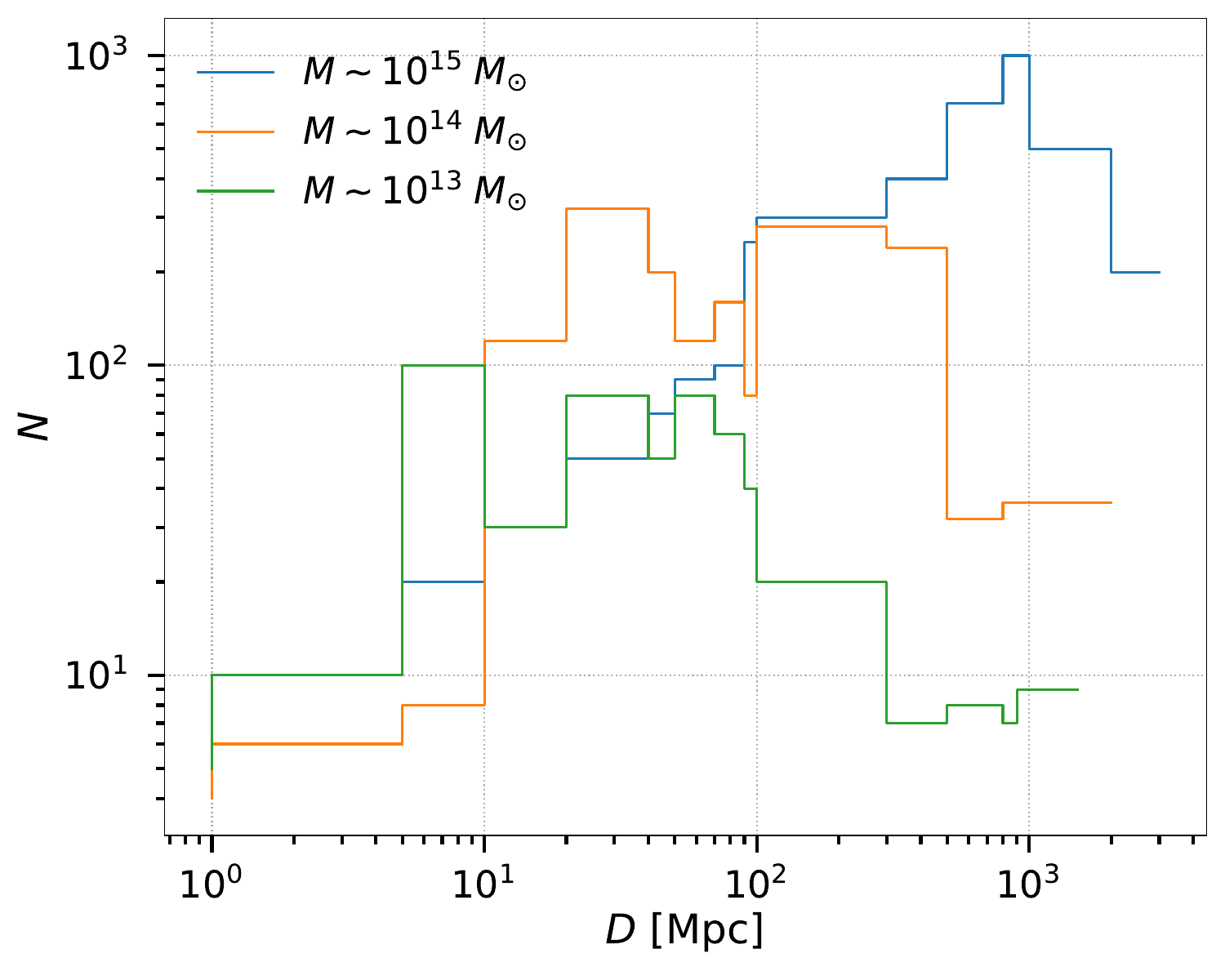}
\caption{\textbf{MFP and trajectory lengths of protons.} Left panel:  MFP  for the processes affecting CRs, namely photopion production, and proton-proton interactions. The target photon fields are the CMB, the EBL (from ref.~\cite{gilmore2012semi}), and the volume-averaged bremsstrahlung radiation for a cluster of mass $M=10^{15} M_\odot$. Right panel: trajectory lengths of CRs emitted at the centre of clusters of mass $10^{13} \text{M}_\odot$, $10^{14} \text{M}_\odot$, and $10^{15} \text{M}_\odot$ with a spectrum $E^{-1}$, for $10^{14} \leq E / \text{eV} \leq 10^{16}$. }
	\label{fig:mfpCR}
\end{figure}

\paragraph{Cosmic-ray propagation in clusters.}
Using the sample of individual clusters obtained from the cosmological simulations, we considered sources of high-energy cosmic rays (CRs) embedded in these structures to compute the associated gamma-ray fluxes. As explained in the Methods of the  main text, we used the CRPropa code~\cite{batista2016crpropa, alvesbatista2022crpropa} for these calculations considering all relevant CR interactions that generate both electrons and photons, namely: photopion production, Bethe-Heitler pair production, and proton-proton ($pp$) interactions. The latter is not natively implemented in the code, so we employed an external CRPropa module described in ref.~\cite{rodriguezramirez2019vhe}, which uses the cross section for $pp$~interactions as parameterised in ref.~\cite{kafexhiu2014parametrization}, given by:
\begin{align} 
\sigma_\text{pp} (E) = \left[ 30.7 - 0.96 \log\left(\dfrac{E}{E_\text{th}}\right) + 0.18 \log^2 \left(\dfrac{E}{E_\text{th}}\right) \right] \times \nonumber \\ 
\left[ 1 - \left(\dfrac{E_\text{th}}{E}\right)^{1.9} \right]^3 \; \text{mb}
\end{align}
with $E$ denoting the kinetic energy for a threshold energy $E_\text{th} \equiv 2 m_\pi c^2 + m^2_\pi c^2 / 2 m_p$. Here $m_\pi$ denotes the mass of the corresponding pion, and $m_p$ the mass of a single proton.
In addition, we have taken into account the energy losses due to the adiabatic expansion of the universe and synchrotron emission. 
We did not make any approximations to describe the properties of the ICM. Instead, we used the background density, temperature, and magnetic fields, directly from the MHD simulations. The temperature distribution in the clusters allowed us to derive the bremsstrahlung radiation field, which we used to extend CRPropa's pre-computed tables of interaction rates to include interactions with the cluster environment. We also used the distributions provided in CRPropa for the other background photon fields, namely the extragalactic background light (EBL)~\cite{dominguez2011extragalactic, gilmore2012semi, stecker2016empirical} and the cosmic microwave background (CMB). The mean free paths (MFPs) for a CR undergoing the aforementioned processes are shown in Fig.~\ref{fig:mfpCR}, left panel. For reference, we also show the expected trajectory length of CRs propagating in individual clusters with different masses (Fig.~\ref{fig:mfpCR}, right panel).

\paragraph{Magnetic confinement of CRs.}
The transport of CRs inside clusters is highly dependent on their masses. The more massive clusters $\gtrsim 10^{14}\; M_{\odot}$ can confine CRs of higher energy for a time longer than the less massive (and smaller) ones. This is consistent with the increase of the particle's Larmor radius with energy while moving inside the cluster. As the cluster mass increases, the transport of CRs changes from diffusive to semi-diffusive or ballistic. This regime change depends on the diffusion coefficient which, in the simplest case, is $D = \langle r^2 \rangle / 6t$, wherein $r$ is the displacement and $t$ the time. Considering a typical size of 1~Mpc and assuming the cluster to exist roughly for a time comparable to the age of the universe, it is possible to obtain an order-of-magnitude estimate of the diffusion coefficient associated to CR confinement/escape, which is $D \sim 10^{27} \; \text{m}^2 \, \text{s}^{-1}$. Comparing this with the Larmor radius of CRs with energy $E \sim 10^{17} \; \text{eV}$, we conclude that in the central regions of the cluster ($r \lesssim 500 \; \text{kpc}$) propagation is diffusion-dominated, whereas in the outskirts CRs can escape the environment. Moreover, high-energy CRs ($E \gtrsim 10^{18}$~eV) propagate (quasi-)balistically. These results agree with those obtained from a simple estimate of the confinement time of a CR, which can be obtained
from the trajectory length ($\ell$): $t \simeq \ell / c \simeq  10^{3}\; \text{Mpc}/c \sim \text{Gyr}$, wherein $c$ denotes the speed of light.
Note that this is comparable with the diffusive escape time found in other works~\cite{inoue2007ultrahigh, batista2018cosmic}. For instance, the acceleration time of a CR up to about  $10^{18}$~eV in a magnetic
field of the order of  $ 10^{-6}$~G for a shock in a cluster is of the order of ~Gyr~\cite{fang2016high}, which is comparable with the diffusive escape time from the acceleration region~\cite{inoue2007ultrahigh}.

\begin{figure}[htb]
\centering
\includegraphics[width=0.9\columnwidth]{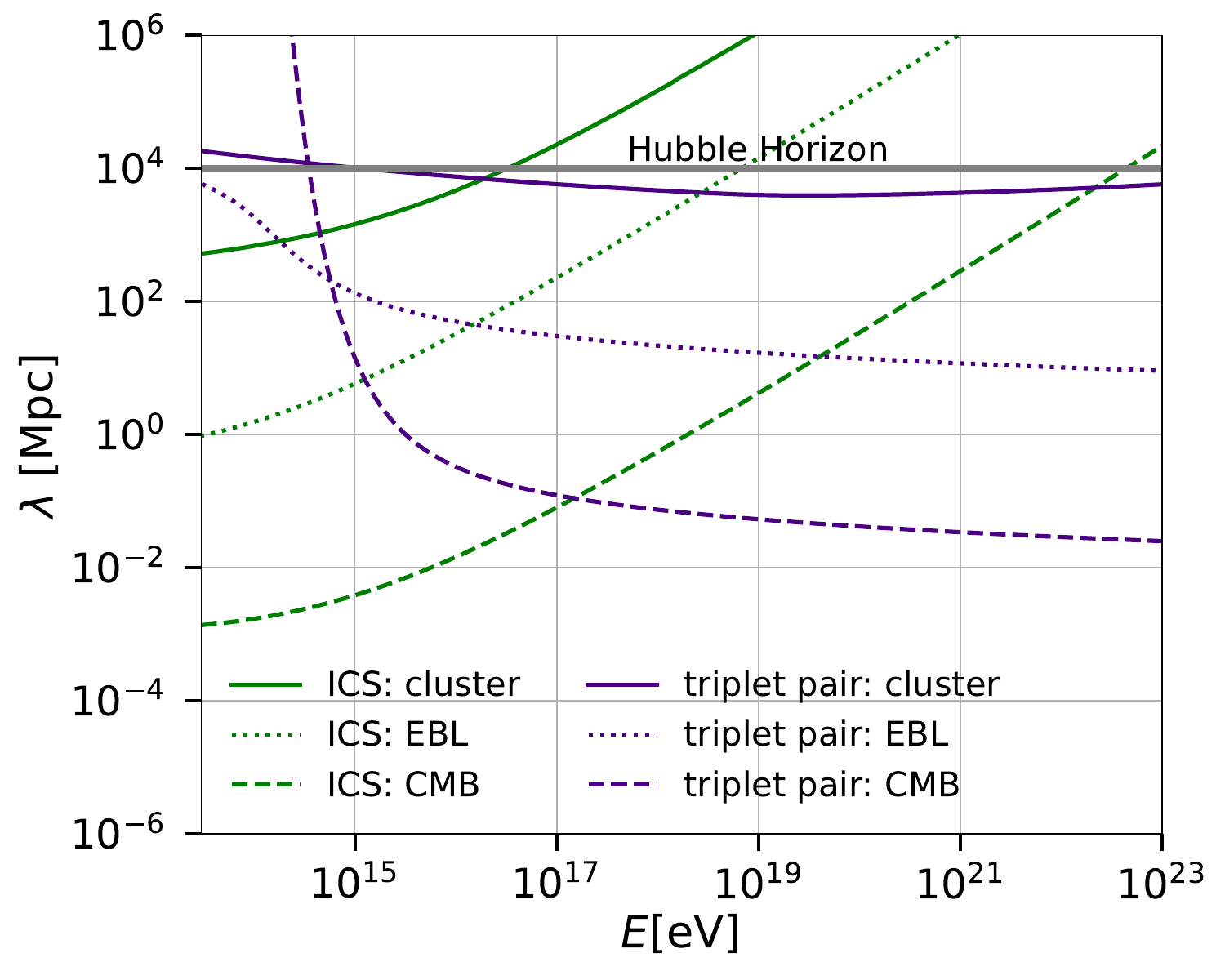}
\includegraphics[width=0.9\columnwidth]{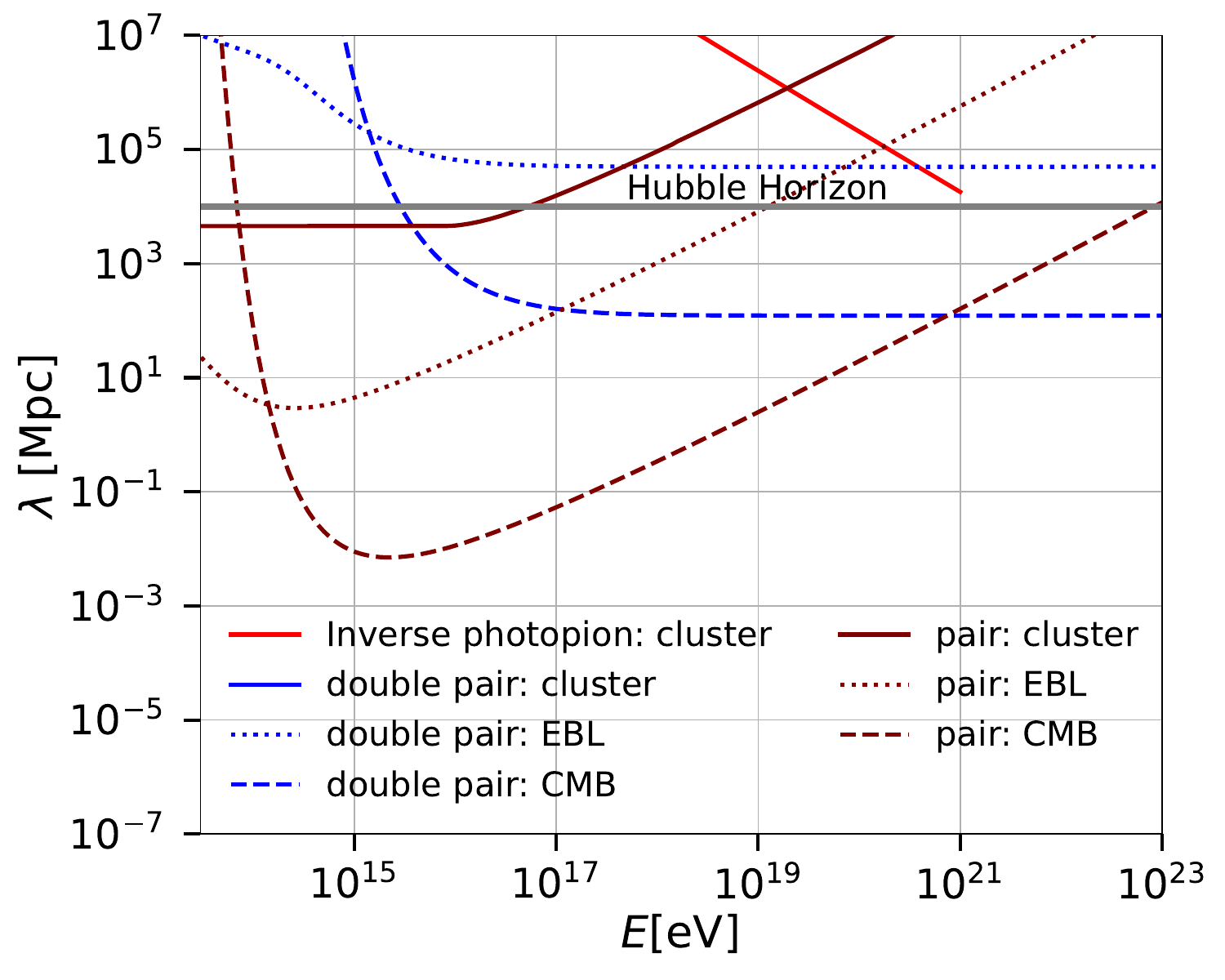}
\caption{\textbf{MFPs of electrons and photons:}
Shown are the processes affecting high-energy electrons (inverse Compton scattering (ICS) and triplet pair production), and photons (pair production and double pair production). The target photon fields are the CMB, the EBL (from ref.~\cite{gilmore2012semi}), and the volume-averaged bremsstrahlung radiation for a cluster of mass $M=10^{15} M_\odot$.}
\label{fig:mfpEM}
\end{figure}

\paragraph{Gamma-ray and electron propagation in clusters.} 
Electrons and photons produced through the processes described in the previous paragraph also undergo interactions, namely: pair production, inverse Compton scattering, double pair production, and triplet pair production. These interactions were taken into account assuming the omnipresent cosmological backgrounds (CMB and EBL), in addition to the ICM photon field due to the bremsstrahlung. The MFPs for these processes are shown in Fig.~\ref{fig:mfpEM} for both electrons (left panel) and photons (right panel). Note that high-energy photons can, in principle, interact with the gas pervading the ICM, which could lead to ``inverse photopion production''. Nevertheless, this channel of interaction is small, as shown on the right panel of Fig.~\ref{fig:mfpEM}, so it was thus neglected.

\paragraph{The effects of CR source evolution.}
The evolutionary history of galaxy clusters implies a constantly changing ICM which, in turn, affects the propagation of CRs and its subsequent gamma-ray emission. For this reason, we have considered a few different scenarios for the evolution of the CR sources embedded in the cluster. This is parametrised through a redshift-dependent function $\psi(z)$. For CR sources following the star-formation rate (SFR), we employ the following expression~\cite{heinze2016cosmogenic, batista2019cosmogenic}:
\begin{equation}\label{eq:SFRz}
\psi_{\text{SFR}}(z) = \frac{1}{B} 
\begin{cases} 
(1+z)^{3.4}  & \text{if} \;\; z \leq 1 \, \\
(1+z)^{-0.3} & \text{if} \;\; 1 < z \leq 4 \, \\
(1+z)^{-3.5} & \text{if} \;\; z > 4 \, . \\
\end{cases}
\end{equation}
Assuming the CR source emissivity is driven by AGNs, the parametrization reads~\cite{heinze2016cosmogenic, batista2019cosmogenic}:
\begin{equation}\label{eq:AGNz}
\psi_{\text{AGN}}(z) = \frac{1}{A} 
\begin{cases}
(1+z)^{5.0} & \text{if} \;\; z \leq 0.97 \, \\
10^{1.09}(1+z)^{1.33} & \text{if} \;\; 0.97 < z \leq 4.48 \, \\
10^{6.66}(1+z)^{-6.2} & \text{if} \;\; z > 4.48 \, . \\
\end{cases}
\end{equation}
Here $A$ and $B$ are normalization constants in Equations~(\ref{eq:AGNz}) 
and~(\ref{eq:SFRz}), respectively.

\paragraph{The effect of star formation and AGN feedback.}
Our simulations are non-radiative and do not include the feedback by active galactic nuclei (AGNs) or star formation, which could reduce the gas density of the clusters and consequently, the gamma-ray flux. To investigate the relevance of these effects to our calculations we followed ref.~\cite{lovisari2015scaling}, which evaluates how a density reduction factor $f(M, z)$ (Equation~$1$ in the main text) can be empirically constrained by observations of different clusters. We find that both effects above produce only minor  modifications in the total flux, as shown in  Figs. 2 and 5 of the main text.

\paragraph{The CR injection spectrum.}
The injection spectrum of CRs ($Q(E)$) is defined as:
\begin{equation}
    Q(E) = \frac{dN}{dE} = Q_0 E^{-\alpha}\exp\left({-\frac{E}{ZR_{\text{max}}}}\right) \,,
\end{equation}
wherein $\alpha$ is the spectral index, $R_\text{max}$ is the maximum rigidity attainable by the CRs, and $Z$ is the atomic number of the CR nucleus ($Z=1$ here). To obtain the normalization constant, $Q_0$, we impose that the total CR energy must be a given fraction ($f_\text{CR}$) of the cluster energy, i.e.:
\begin{equation}
    \int\limits_{E_\text{min}}^{E_\text{max}} \text{d}E \; E Q(E) = f_\text{CR} E_\text{tot} \,,
    \label{eq:normalisation}
\end{equation}
where $E_\text{tot}$ refers to the total energy of the cluster corresponding to a luminosity $L_\text{tot}$. Here the minimum and maximum energies, $E_\text{min} \simeq 1 \; \text{GeV}$ and $E_\text{max} \simeq 10 \; \text{EeV}$, are essentially the rest mass of the CR and the maximum energy a CR could reach according to the model, respectively. The later, in particular, was conservatively chosen to be $10 \; \text{EeV}$ because already at much lower energies (about  $ 0.1 \; \text{EeV}$) CRs can escape clusters without effectively interacting with the ICM.

\paragraph{A note on the composition of the CRs.}
We assumed only proton composition of CRs, because we expect a much smaller contribution from heavier elements  (see, e.g., \cite{kotera2009propagation}), especially if they are produced in large-scale shocks. If there are heavier nuclei in clusters, they should be subdominant with respect to protons because CR acceleration depends on rigidity (energy over charge). Nevertheless, CR sources within clusters such as starburst galaxies have high supernova rates, and compact objects such as magnetars wherein heavier energetic CRs can be accelerated, such that the gamma-ray flux may change.  
It is also worth mentioning that gamma rays (and neutrinos) in general tend to be produced more through processes involving protons than nuclei. This is because hadronic gamma rays are created mostly through the decay of pions, and in the case of heavier CR nuclei, photodisintegration tends to dominate over pion-producing mechanisms. 
Furthermore, photons produced, for instance, by electron/positrons generated via nuclear beta decays (in the photodisintegration chain, for example) are generally not sufficient to lead to appreciable fluxes of high-energy photons. 

The injected energy range of CRs is $10^{11} \leq E/\text{eV} \leq10^{19}$, which leads to a peak of the integrated gamma-ray flux at energies around $10$~GeV as shown in Figs. 2 and 4  of the main text (see also Fig. \ref{fig:ph_UDSSMP}). 

\begin{figure}[htb!]
\centering
\includegraphics[width=0.8\columnwidth]{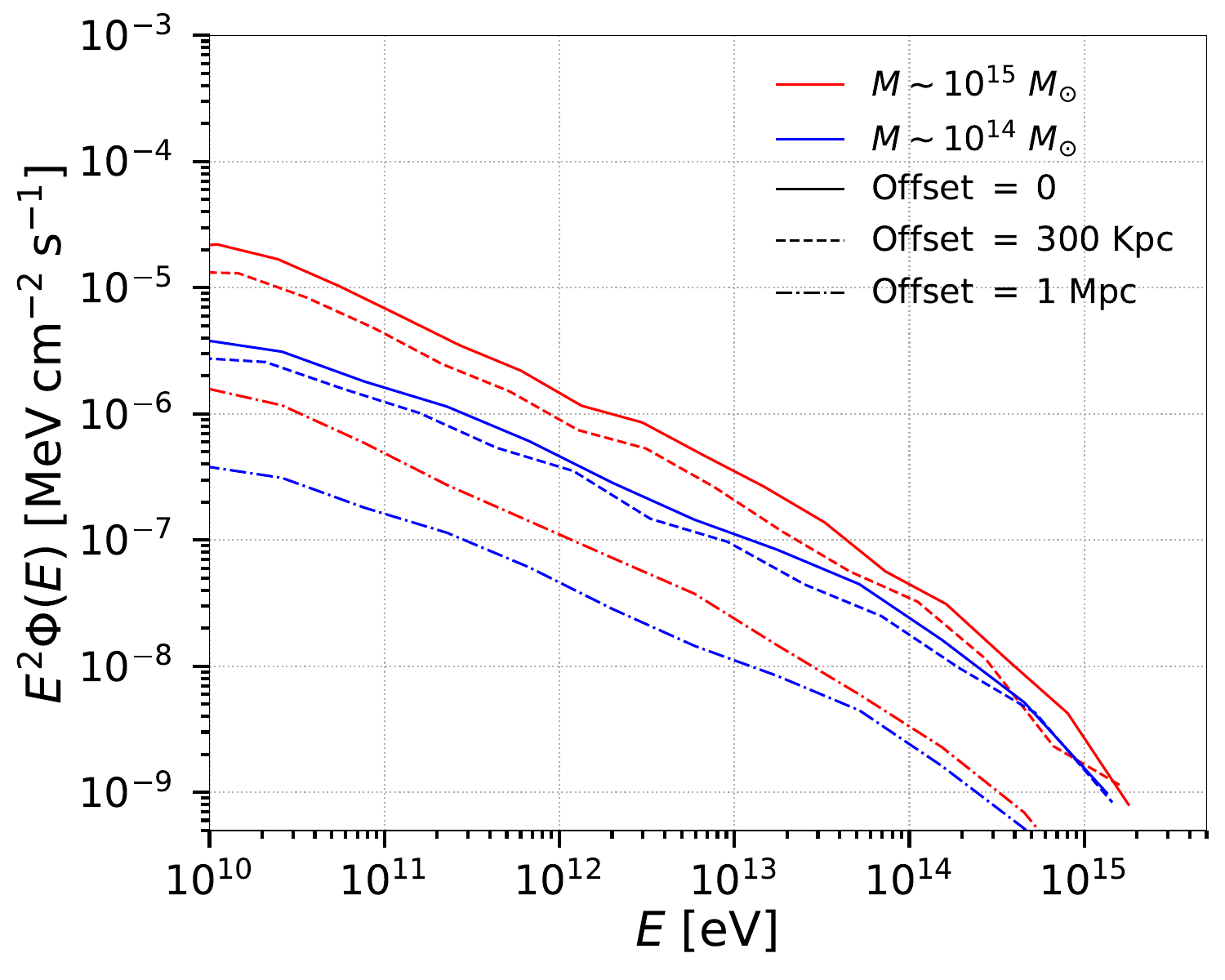}
\caption{
\textbf{Gamma-ray flux at the edge of individual clusters} (radius $\sim 2$~Mpc) of mass $M\sim 10^{15} \; M_{\odot}$ (red); and $10^{14}\; M_{\odot}$ (green), at redshift $z\sim 0$. We considered CR  sources  located at the center of the cluster (solid lines), at  $300$ kpc (dashed lines), and at $1$ Mpc away from the center (dash-dotted lines). The spectral index of the CR spectrum has a power-law index $\alpha=2.3$ and an exponential energy cut-off $ E_{\text{max}} = 10^{17}$~eV.
}
\label{fig:ph_offset}
\end{figure}

\paragraph{The gamma-ray flux from individual clusters.} Our analysis involve computing first the gamma-ray flux emitted by individual clusters. This is shown in Fig.~\ref{fig:ph_offset} for masses $M\sim 10^{15} \; M_{\odot}$ and $10^{14}\; M_{\odot}$. The figure also shows the dependence of the photon flux on the position of the CR sources inside the clusters. As expected, the photon production rate is smaller when the source is located farther away from the centre.
Note that the central regions of clusters are more densely populated than the outskirts. Therefore, it is a reasonable approximation to consider all CR sources inside clusters to be at their centres, since the contribution of marginal sources is lower by nearly a ten-fold, as shown in Fig.~\ref{fig:ph_offset}. This implies that even under these assumptions the total gamma-ray flux would be overestimated by less than an order of magnitude.

A sanity check for our calculations is to compare the results for a few individual clusters selected from our simulations with observations. Fermi-LAT~\cite{ackermann2014search}, for instance, obtained upper limits for the emission from three Abell clusters, A400, A1367, and A3112. At $E \sim 10 \; \text{GeV}$, the most stringent amongst these limits is $\simeq 4.4 \times 10^{-7} \; \text{MeV} \, \text{cm}^{-2} \, \text{s}^{-1}$. Although Fig.~\ref{fig:ph_offset} is for the flux at the edge of the cluster, without intergalactic propagation, its total gamma-ray energy is $E_\text{tot}^\text{sim} \lesssim 2 \times 10^{43} \; \text{MeV}$, which is much less than the total energy inferred from observations, $E_\text{tot}^\text{obs} \sim 4 \times 10^{45} \; \text{MeV}$, considering the cluster A400, distant approximately $100 \; \text{Mpc}$ from Earth. There is also a geometrical correction factor due to the fact that the simulated flux is divided by a solid angle, but this should not exceed one order of magnitude in the most conservative case. Therefore, this simple ballpark estimate confirms that our gamma-ray estimates for individual clusters are safely compatible with present-day observational constraints~\cite{ackermann2014search, magic2016deep,  fermi2016search}.

\begin{figure}[htb!]
\centering
\includegraphics[width=0.9\columnwidth]{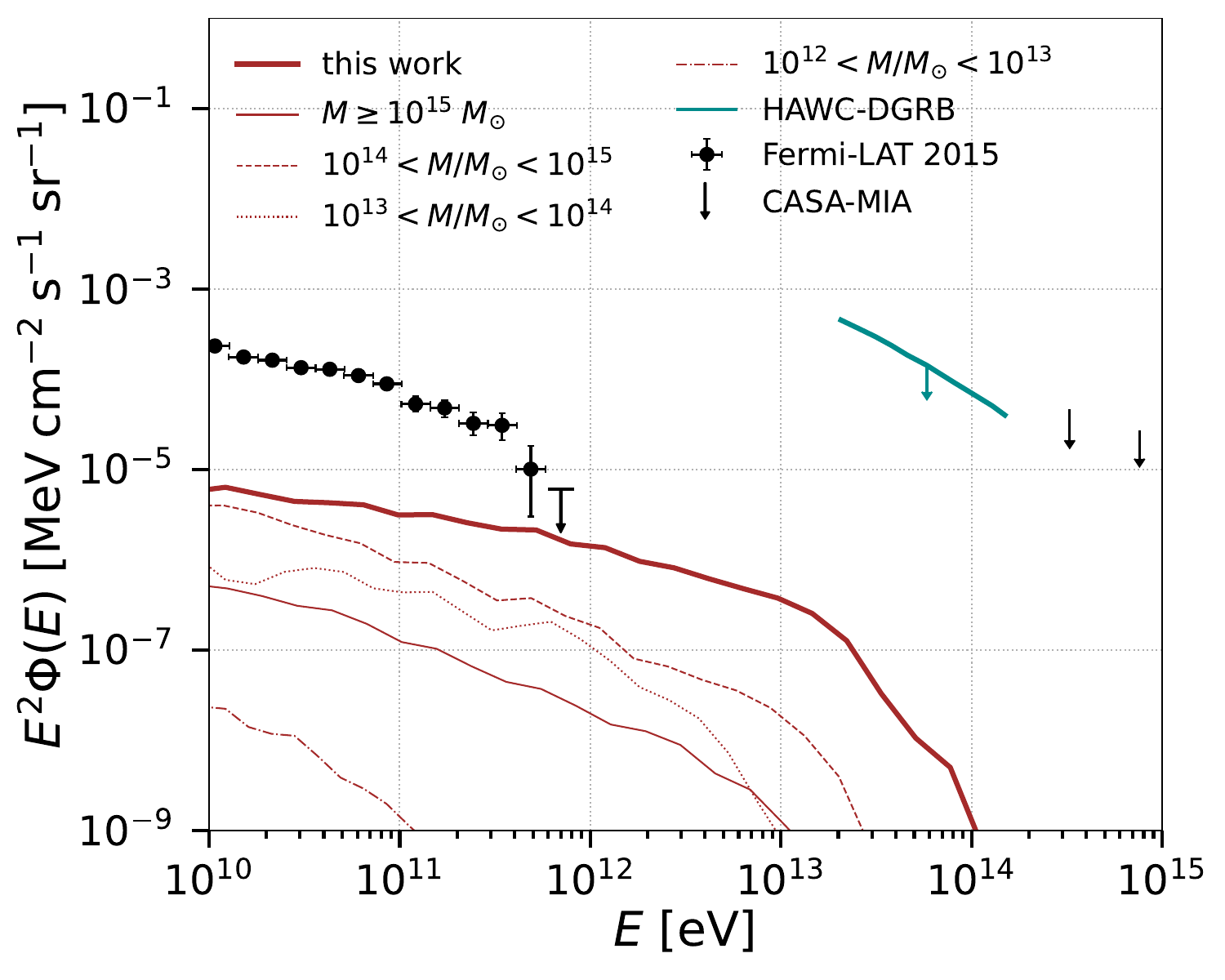}
\caption{\textbf{Contribution to the integrated  gamma-ray flux from different cluster mass ranges.} The flux is plotted for CR spectral parameters $\alpha = 2.3$  and $E_\text{max} = 10^{17}$~eV. 
Fermi-LAT data for the DGRB (error bars correspond to the total uncertainties, statistical and systematic) \cite{ackermann2015spectrum}, upper limits from  HAWC ($95\%$ confidence level)~\cite{harding2019constraints} and CASA-MIA ($90\%$ confidence level) \cite{chantell1997limits} are also shown for comparison.
}
\label{fig:Gamma_massRang}
\end{figure}

\paragraph{The integrated gamma-ray flux from different  cluster mass ranges.}
In Fig.~\ref{fig:Gamma_massRang}, we present the dependence of the gamma-ray flux on the mass of the clusters. The major contribution comes from  clusters in the mass range $10^{13} \lesssim M/M_\odot \lesssim 10^{15}$.

\begin{figure}[htb!]
\centering 
\includegraphics[width=0.9\columnwidth]{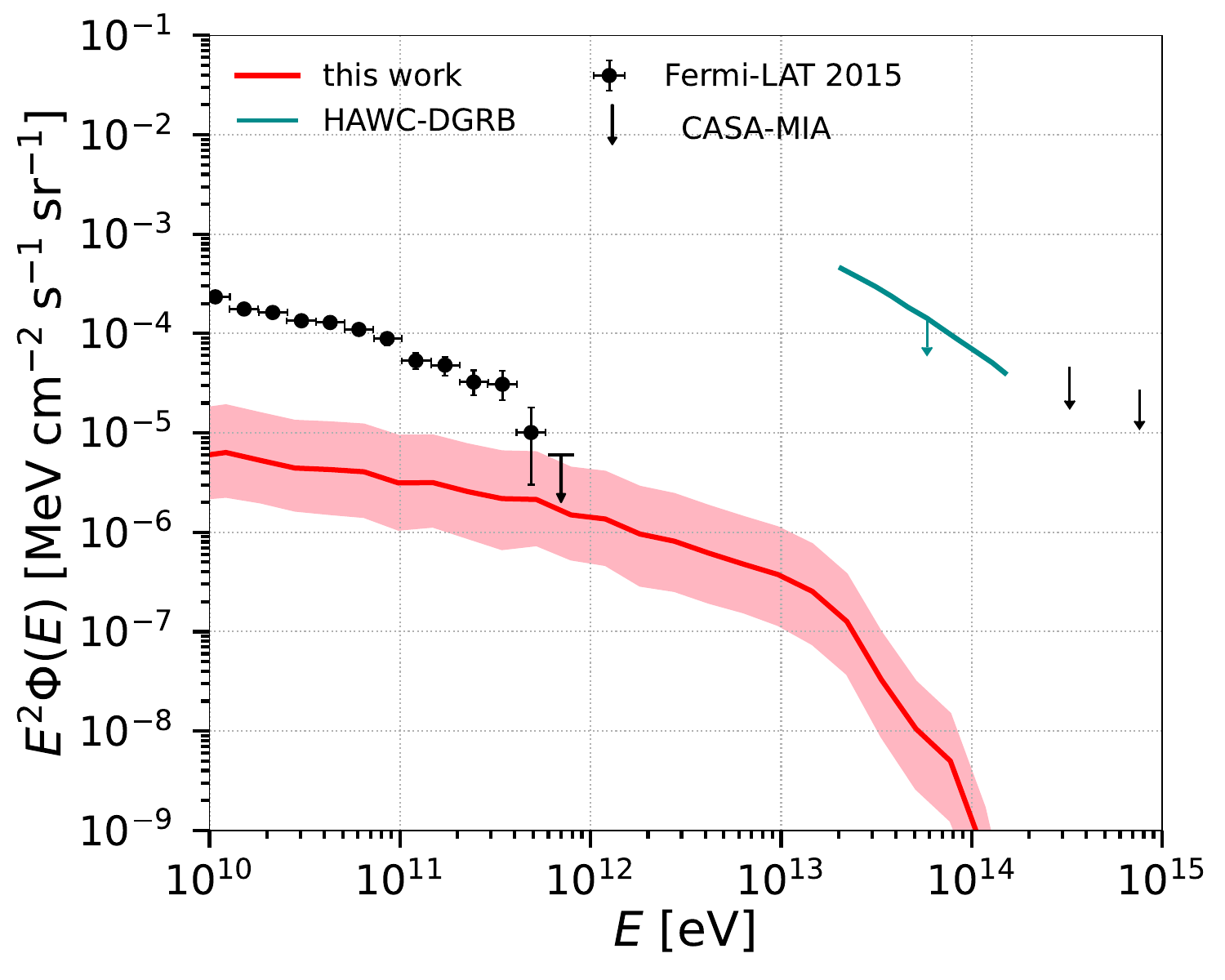}
\caption{\textbf{Integrated gamma-ray flux from clusters for different values of the fraction of the cluster luminosity that goes into CRs.} The pink band corresponds to $(0.5 - 5)\%$ of this fraction and the red line corresponds to $1\%$. Here we assume a CR spectral index $\alpha=2.3$ and a cutoff energy $E_\text{max}= 10^{17}$ eV. Fermi-LAT data for the DGRB (error bars correspond to the total uncertainties, statistical and systematic) \cite{ackermann2015spectrum}, 
upper limits from  HAWC ($95\%$ confidence level)~\cite{harding2019constraints} and CASA-MIA ($90\%$ confidence level) \cite{chantell1997limits} are also shown.
}\label{fig:CRpower}
\end{figure}

\begin{figure}[htb!]
\centering
\includegraphics[width=1.0\columnwidth]{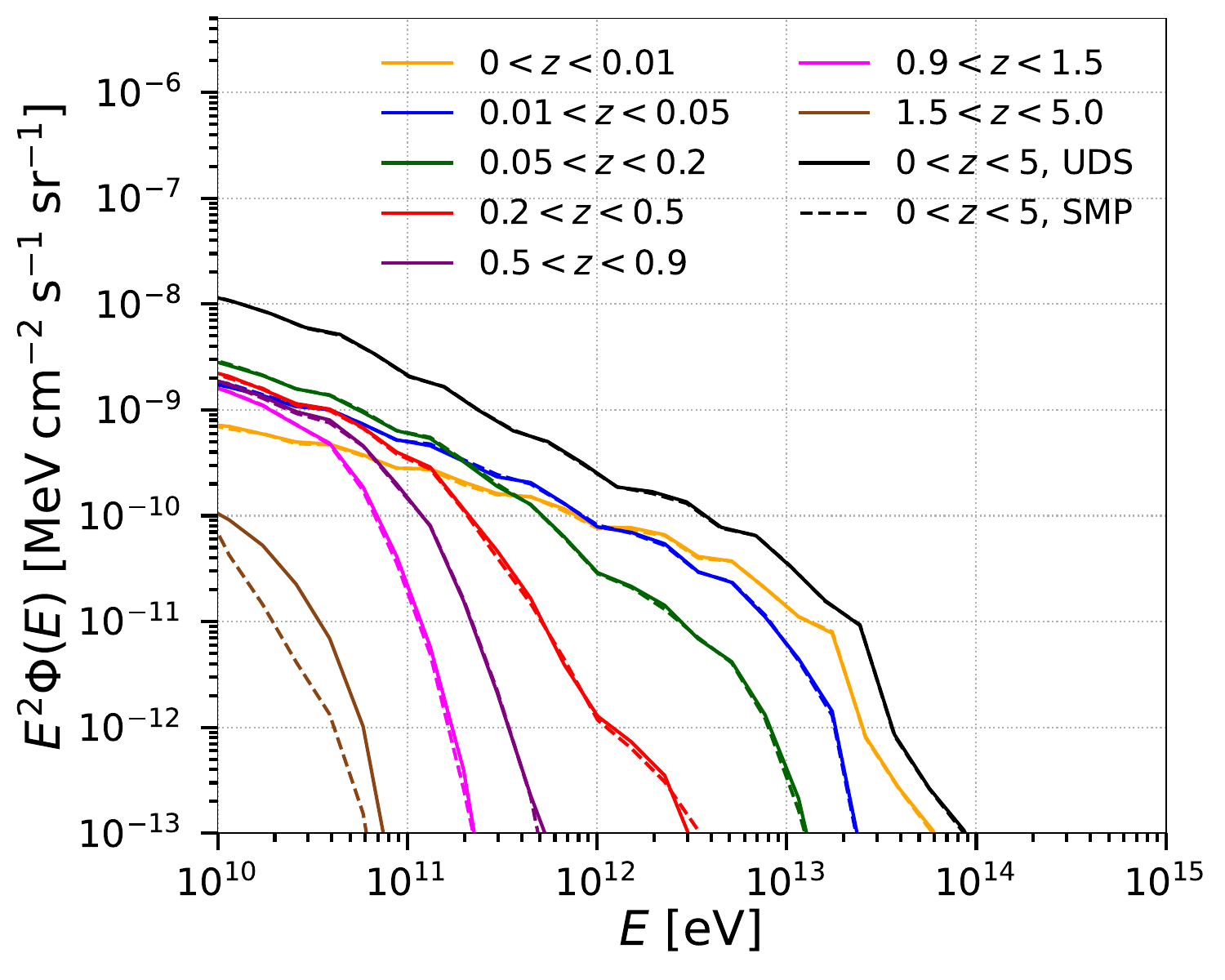}
\caption{\textbf{Flux of uniform distribution of CR sources (UDS) vs randomly distributed sources (SMP1)}, UDS and SMP1 are represented by solid and dashed lines, respectively.
The spectral index and cutoff energy is $\alpha = 2.3$ and $E_{\text{max}}=10^{17}$ eV, respectively. This figure is for the EBL model of ref.~\cite{gilmore2012semi}. }
\label{fig:ph_UDSSMP}
\end{figure}

\paragraph{The CR luminosity.} 
Throughout this work, in order to compute the gamma-ray flux, we have considered that $f_\text{CR} \sim 1\%$ of the luminosity of a cluster goes into CRs, which is consistent with estimates from observations (see e.g. refs.~\cite{ackermann2014search, pinzke2010simulating}), as stressed in the main text. However, to illustrate the relevance of this parameter, in Fig.~\ref{fig:CRpower}, we show the gamma-ray flux spanning a range of values of this fraction, namely, $f_\text{CR} \sim (0.5 - 5)\%$ of the luminosity of the clusters going into CRs. The results indicate that the variation is not substantial, i.e., it is at most of one order of magnitude, which is compatible with the linear dependence between flux and luminosity.

\paragraph{Integrated gamma-ray flux for different redshift intervals.}
As indicated in Fig. 2 of the main text, the major contribution to the integrated  flux comes from CR sources at low redshifts $z \lesssim 0.3$, whose flux is less attenuated by interactions with the EBL.
This suggests that  the resulting spectral hardening is due to this low redshift contribution mostly. This result is reassured by Fig. \ref{fig:ph_UDSSMP}, which shows the gamma-ray flux for all redshift intervals, and is complementary to Fig. 2 of the main text.

\begin{figure*}[htb!]
\centering
\includegraphics[width=1.5\columnwidth]{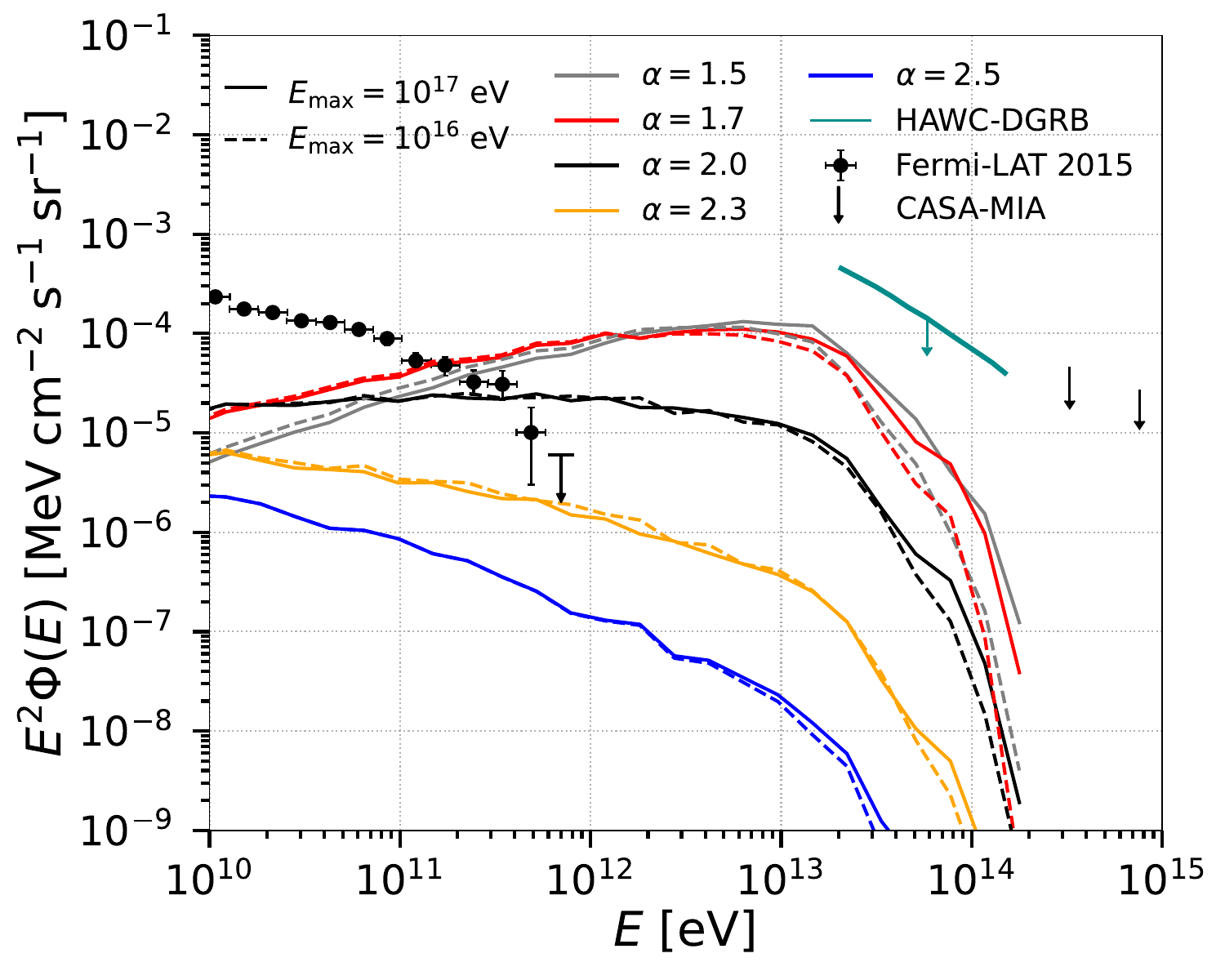} 
\caption{\textbf{Integrated flux for different combinations of $\alpha \; \text{and} \; E_\text{max}$} and comparison with the Fermi-LAT data (error bars correspond to the total uncertainties, statistical and systematic)
) \cite{ackermann2015spectrum}, and CASA-MIA ($90\%$ confidence level) \cite{chantell1997limits} and HAWC DGRB ($95\%$ confidence level) upper limits \cite{harding2019constraints}. }
\label{fig:ph_AsEs_SrcEv}
\end{figure*}

 \paragraph{Integrated gamma-ray flux for different CR spectral parameters.}
 In Fig.~\ref{fig:ph_AsEs_SrcEv} we show the gamma-ray flux for different combinations of the parameters $\alpha \; \text{and} \; E_{\text{max}}$ of the CRs. The choice of this parametric range is discussed in detail in the main text.

\begin{figure*}[htb!]
\centering 
\includegraphics[width=1.5\columnwidth]{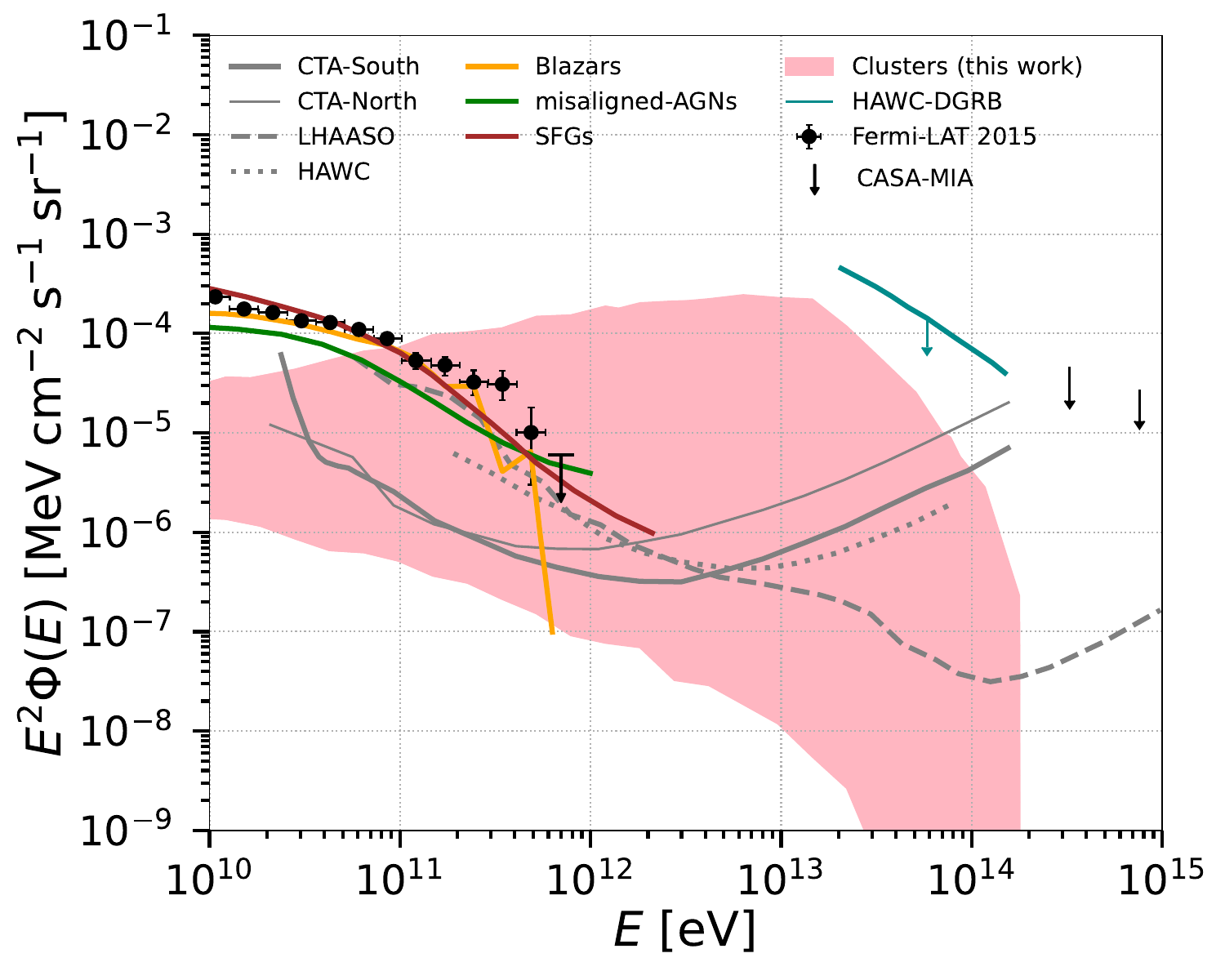}
\caption{\textbf{Contribution to DGRB from different types of astrophysical sources.} The pink band is plotted for the fiducial range of parameters in our work $\alpha = 1.5 - 2.5, \; E_\text{max} = 10^{16} - 10^{17}\; \text{eV}$.  Besides showing the observed DGRB flux from Fermi-LAT (error bars correspond to the total uncertainties, statistical and systematic) \cite{ackermann2015spectrum} and upper limits from  HAWC ($95\%$ confidence level)~\cite{harding2019constraints} and CASA-MIA ($90\%$ confidence level) \cite{chantell1997limits}, this figure also presents the sensitivity curves obtained for point sources from LHAASO \cite{di2016lhaaso}, HAWC \cite{abeysekara2013sensitivity}, and the forthcoming CTA North and South observatories \cite{cta2018science} for comparison (gray curves).
These sensitivity curves are shown only for reference and the scaling factor is simply $\sim \; \text{PSF}^2/4\pi$, where PSF is the point spread function.
We also show the contribution from individual sources to the DGRB, namely,  blazars \cite{ajello2015origin}, AGNs \cite{di2013diffuse}, and SFGs \cite{roth2021diffuse}.
}\label{fig:DGB_AllSrc}
\end{figure*}

\paragraph{Integrated gamma-ray flux for entire parametric space.}
As stressed previously, the CRs are injected with a minimum energy of $100$~GeV, and our analysis of the gamma-ray flux  produced by them in the clusters extends down to  $10$~GeV approximately.
In any case, our main interest is the contribution of the clusters to the higher energies of the DGRB, whose origin is more uncertain, less constrained and, in principle, not explained by point sources or individual source populations, as shown in Fig.~\ref{fig:DGB_AllSrc}.
We have also plotted in this figure point-like sensitivity curves of different gamma-ray observatories, which were rescaled by an appropriate angular factor. This is meant to be a reference only. 
There are experimental difficulties in measuring an all-sky flux with  
Cherenkov telescopes 
with relatively small fields of view such as CTA. The message intended is: if CTA could scan the whole sky and measure a diffuse flux of gamma rays,
then the ideal curve obtained by a direct scaling of the point-source sensitivity would be the one shown.
In fact, this figure evidences that the major contribution to the DGRB below about  $ 400$~GeV most probably comes from  individual sources \cite{ackermann2016resolving} such as blazars \cite{ajello2015origin}, AGNs~\cite{di2013diffuse}, and SFGs~\cite{roth2021diffuse}.
But for energies greater than $\gtrsim 100$~GeV, our simulations indicate that galaxy clusters can also contribute substantially to the DGRB. 
This contribution could amount to up to $100\%$ observed flux by Fermi-LAT, for spectral indices $\alpha \leq 2$ and maximum energies $E_\text{max} \geq 10^{17}$~eV.

\paragraph{Remarks on the interpretation of the results.} Our goal in this work was not to fit the data observed by Fermi-LAT. Instead, we calculated the high-energy gamma-ray flux that can be produced by the entire galaxy cluster population, considering a reasonable  set of free parameters, and performing the most detailed treatment so far employing 3D simulations (in contrast, many studies until now adopted a semi-analytic approach and/or simplified 1D calculations). Uncertainties in such determinations are always expected. The question is whether they lead to order-of-magnitude changes in the results. To answer this question, we explored the parametric space of the potentially most influential quantities, i.e., we considered a fiducial range of the CR spectral parameters, which are all compatible with theoretical/observational expectations. Moreover, we employed a detailed treatment of intergalactic gamma-ray propagation, including EBL uncertainties (see Fig. 3 of the main text). 
Fermi observations and upper limits obtained from HAWC and CASA-MIA for the DGRB, depicted in Fig. 5  of the main text, and  Fig. \ref{fig:ph_AsEs_SrcEv}, clearly put constraints on the parametric space we swept.
Our results turn out to be compatible with these  constraints for spectral indices $\gtrsim 2.3$, considering our fiducial parameters.
Therefore, though uncertainties remain, such as the determination of the effects of the still-unknown magnetic fields of the diffuse IGM on the gamma-ray cascading after emerging from individual clusters (as described in the main text), we believe we have covered most of the fundamental parametric space, thus constraining the uncertainties in the flux to less than one order of magnitude.

\end{document}